\documentclass[10pt]{article}
\usepackage{amsmath}
\usepackage{amsfonts}
\usepackage{graphicx}
\usepackage{graphics}
\usepackage{theorem}
\usepackage{color}

\usepackage{amsxtra}
\usepackage{amstext}
\usepackage{amssymb}
\usepackage{latexsym}

\usepackage[square]{natbib}
\usepackage{graphicx}
\usepackage{dcolumn}
\usepackage{bm}

\makeatletter
\newcommand{\cn}{\mathop{\operator@font cn}}
\makeatother
\makeatletter
\newcommand{\sn}{\mathop{\operator@font sn}}
\makeatother
\makeatletter
\newcommand{\dn}{\mathop{\operator@font dn}}
\makeatother
\newcommand{\Lame} {Lam\'{e} }
\newcommand{\Saito} {Sait\^{o}}
\newcommand{\Four} {Fourier }
\newcommand{\Schr} {Schr\"{o}dinger }

\begin{document}


\nocite{*}


\title{Chaotic Dynamics of N--degree of Freedom Hamiltonian Systems}
\author{Chris ANTONOPOULOS\footnote{E--mail: {\tt antonop@math.upatras.gr}}$^{\ 1}$,
Tassos BOUNTIS\footnote{E--mail: {\tt bountis@math.upatras.gr}}$^{\
1}$ \\ and \\ Charalampos SKOKOS\footnote{E--mail: {\tt
hskokos@cc.uoa.gr}}$^{\ 123}$}
\date{\today}
\maketitle
\begin{center}
{\it $^1$Department of Mathematics\\and\\Center for Research and
Applications of Nonlinear Systems (CRANS),\\University of
Patras,\\ GR--$26500$, Rio, Patras, Greece\\
\vspace{0.5cm}$^2$Research Center for Astronomy and Applied
Mathematics\\Academy
of Athens,\\Soranou Efesiou $4$, GR--$11527$, Athens, Greece\\
\vspace{0.5cm}$^3$Department of Applications of Informatics
and Management in Finance,\\
Technological Institute of Mesologhi,\\GR--$30200$, Mesologhi,
Greece}
\end{center}
\begin{abstract}
We investigate the connection between local and global dynamics of
two $N$--degree of freedom Hamiltonian systems with different
origins describing one--dimensional nonlinear lattices: The
Fermi--Pasta--Ulam (FP\-U) model and a discretized version of the
nonlinear \Schr equation related to the Bose--Einstein Condensation
(BEC). We study solutions starting in the immediate vicinity of
simple periodic orbits (SPOs) representing in--phase (IPM) and
out--of--phase motion (OPM), which are known in closed form and
whose linear stability can be analyzed exactly. Our results verify
that as the energy $E$ increases for fixed $N$, beyond the
destabilization threshold of these orbits, all positive Lyapunov
exponents $L_i,\ i=1,\ldots,N-1$, exhibit a transition between two
power laws, $L_i\propto E^{B_k}$, $B_k>0,\ k=1,2$, occurring at the
same value of $E$. The destabilization energy $E_c$ per particle
goes to zero as $N\rightarrow\infty$ following a simple power--law,
$E_c/N\propto N^{-\alpha}$, with $\alpha$ being 1 or 2 for the cases
we studied. However, using the SALI, a very efficient indicator we
have recently introduced for distinguishing order from chaos, we
find that the two Hamiltonians have very different dynamics near
their {\it stable} SPOs: For example, in the case of the FPU system,
as the energy increases for fixed $N$, the islands of stability
around the OPM decrease in size, the orbit destabilizes through
period--doubling bifurcation and its eigenvalues move steadily away
from $-1$, while for the BEC model the OPM has islands around it
which {\it grow} in size before it bifurcates through symmetry
breaking, while its real eigenvalues return to $+1$ at very high
energies. Furthermore, the IPM orbit of the BEC Hamiltonian {\it
never destabilizes}, having finite--sized islands around it, even
for very high $N$ and $E$. Still, when calculating Lyapunov spectra,
we find for the OPMs of both Hamiltonians that the Lyapunov
exponents decrease following an exponential law and yield extensive
Kolmogorov--Sinai entropies per particle
$h_{KS}/N\propto\mathrm{const.}$, in the thermodynamic limit of
fixed energy density $E/N$ with $E$ and $N$ arbitrarily large.
\end{abstract}
\section{Introduction}\label{intro}
Chaotic behavior in Hamiltonian systems with many degrees of
freedom has been the subject of intense investigation in the last
fifty years, see e.g. [Lichtenberg \& Lieberman{,} 1991; MacKay \&
Meiss{,} 1987; Wiggins{,} 1988] and \cite{Simo}. By degrees of
freedom (dof) we are referring to the number of canonically
conjugate pairs of positions and momentum variables, $q_k$ and
$p_k$ respectively, with $k=1,2,\ldots,N$. The relevance of these
systems to problems of practical concern cannot be overemphasized.
Their applications range from the stability of the solar system
\cite{Contopoulos} and the containment of charged particles in
high intensity magnetic fields \cite{Lichtenberg} to the blow--up
of hadron beams in high energy accelerators \cite{Scandale} and
the understanding of the properties of simple molecules and
hydrogen--bonded systems \cite{Bountis_1992, Prosmiti}.

One of the most fundamental areas in which the dynamics of
multi--degree of freedom Hamiltonian systems has played (and
continues to play) a crucial role is the study of transport
phenomena in one--dimensional ($1$D) lattices and the role of
chaos in providing a link between deterministic and statistical
behavior \cite{Chirikov,Lichtenberg,Ford}. In this context, a
lattice of $N$ dof is expected, in the thermodynamic limit
($N\rightarrow\infty$ at fixed energy density $E/N$) to exhibit
chaotic behavior for almost all initial configurations, satisfying
at least the property of ergodicity. This would allow the use of
probability densities, leading from the computation of orbits to
the study of statistical quantities like ensemble averages and
transport coefficients.

Chaotic regions, where nearby solutions diverge exponentially from
each other, provide an excellent ``stage'' on which such a desired
transition from classical to statistical mechanics can occur.
However, the presence of significantly sized islands (or tori) of
quasiperiodic motion, in which the dynamics is ``stable'' for long
times, preclude the success of this scenario and make any attempt at
a globally valid statistical description seriously questionable.
These tori occur e.g. around {\it simple periodic orbits} which are
stable under small perturbations and, if their size does not shrink
to zero as $N$ or $E$ increases, their presence can attribute global
consequences to a truly local phenomenon.

That such important periodic orbits do exist in Hamiltonian
lattices, even in the $N\rightarrow\infty$ limit, could not have
been more dramatically manifested than in the remarkable discovery
of discrete breathers (see e.g. \cite{Flach}), which by now have
been observed in a great many experimental situations [Eisenberg
{\it et al.}, 1998; Schwarz {\it et al.}, 1999; Fleischer {\it et
al.}, 2003; Sato {\it et al.}, 2003]. Discrete breathers are
precisely one such kind of stable periodic orbits, which also
happen to be localized in space, thus representing a very serious
limitation to energy transport in nonlinear lattices.

Then, there were, of course, the famous numerical experiments of
Fermi, Pasta and Ulam of the middle $1950$'s \cite{Fermi_1955},
which demonstrated the existence of recurrences that prevent energy
equipartition among the modes of certain $1$D lattices, containing
nonlinear interactions between nearest neighbors. These finite,
so--called FPU Hamiltonian systems were later shown to exhibit a
transition to ``global'' chaos, at high enough energies where major
resonances overlap \cite{Izrailev}. Before that transition, however,
an energy threshold to a ``weak'' form of chaos was later discovered
that relies on the interaction of the first few lowest frequency
modes and, at least for the FPU system, does appear to ensure
equipartition among all modes [De Luca {\it et al.}, $1995$; De Luca
\& Lichtenberg, $2002$]. Interestingly enough, very recently, this
transition to ``weak'' chaos was shown to be closely related to the
destabilization of one of the lowest frequency nonlinear normal mode
of this FPU system \cite{Flach_2005}. Thus, today, $50$ years after
its famous discovery, the Fermi--Pasta--Ulam problem and its
transition from recurrences to true statistical behavior is still a
subject of ongoing investigation \cite{Berman}.

In this paper, we have sought to approach the problem of global
chaos in Hamiltonian systems, by considering two paradigms of $N$
dof, $1$D nonlinear lattices, with very different origins.

One is the famous FPU lattice mentioned above, with quadratic and
quartic nearest neighbor interactions, described by the
Hamiltonian
\begin{equation}\label{FPU_Hamiltonian}
H=\frac{1}{2}\sum_{j=1}^{N}\dot{x}_{j}^{2}+\sum_{j=0}^{N}\biggl
(\frac{1}{2}(x_{j+1}-x_{j})^2+\frac{1}{4}\beta(x_{j+1}-x_{j})^4\biggr)=E
\end{equation}
where $x_{j}$ is the displacement of the $j$th particle from its
equilibrium position, $\dot{x}_{j}$ is the corresponding
canonically conjugate momentum of $x_{j}$, $\beta$ is a positive
real constant and $E$ is the value of the Hamiltonian representing
the total energy of the system.

The other one is obtained by a discretization of a partial
differential equation (PDE) of the nonlinear \Schr type referred
to as the Gross--Pitaevskii equation \cite{Dalfovo}, which in
dimensionless form reads
\begin{equation}\label{Gross--Pitaevskii equation}
i\hslash\frac{\partial\Psi(x,t)}{\partial
t}=-\frac{\hslash^{2}}{2}\frac{\partial^{2}\Psi(x,t)}{\partial
x^2}+V(x)\Psi(x,t)+g|\Psi(x,t)|^{2}\Psi(x,t),\ i^2=-1
\end{equation}
where $\hslash$ is the Planck constant, $g$ is a positive constant
(repulsive interactions between atoms in the condensate) and $V(x)$
is an external potential.

Equation (\ref{Gross--Pitaevskii equation}) is related to the
phenomenon of Bose--Einstein Condensation (BEC) \cite{Ketterle}.
Here we consider the simple case $V(x)=0$, $\hslash=1$ and
discretize the $x$--dependence of the complex variable
$\Psi(x,t)\equiv\Psi_j(t)$ in (\ref{Gross--Pitaevskii equation}),
approximating the second order derivative by
$\Psi_{xx}\simeq\frac{\Psi_{j+1} + \Psi_{j-1} -2\Psi_{j}}{\delta
x^2}$. Setting then $\Psi_{j}(t) = q_{j}(t)+ i\cdot p_{j}(t),\
i^2=-1,\;j=1,2,\ldots,N$ and $|\Psi(x,t)|^{2}=q_j^2(t)+p_j^2(t)$,
one immediately obtains from the above PDE (\ref{Gross--Pitaevskii
equation}) a set of ordinary differential equations (ODEs) for the
canonically conjugate variables, $p_j$ and $q_j$, described by the
BEC Hamiltonian \cite{Trombettoni_2001,Smerzi_2003}
\begin{equation}\label{BEC_Hamiltonian}
H=\frac{1}{2}\sum_{j=1}^{N}(p_{j}^{2}+q_{j}^{2})+\frac{\gamma}{8}
\sum_{j=1}^{N}(p_{j}^{2}+q_{j}^{2})^2-\frac{\epsilon}{2}\sum_{j=1}^{N}
(p_{j}p_{j+1}+q_{j}q_{j+1})=E
\end{equation}
where $\gamma>0$ and $\epsilon=1$ are constant parameters,
$g=\frac{\gamma}{2}>0$ with $\delta x=1$ and $E$ is the total
energy of the system.

In Sec.~\ref{sec_2}, we study these Hamiltonians, focusing on some
simple periodic orbits (SPOs), which are known in closed form and
whose local (linear) stability analysis can be carried out to
arbitrary accuracy. By SPOs, we refer here to periodic solutions
where all variables return to their initial state after only one
maximum and one minimum in their oscillation, i.e. all
characteristic frequencies have unit ratios. In particular, we
examine first their bifurcation properties to determine whether
they remain stable for arbitrarily large $E$ and $N$, having
perhaps finitely sized islands of regular motion around them. This
was found to be true only for the so--called in--phase--mode (IPM)
of the BEC Hamiltonian (\ref{BEC_Hamiltonian}).

SPOs corresponding to out--of--phase motion (OPM) of either the FPU
(\ref{FPU_Hamiltonian}) or the BEC system (\ref{BEC_Hamiltonian})
destabilize at energy densities $\frac{E_c}{N}\propto N^{-\alpha}$,
with $\alpha=1$ or $2$ (for the SPOs we studied), as
$N\rightarrow\infty$. The same result was also obtained for what we
call the OHS mode of the FPU system \cite{Ooyama}, where all even
indexed particles are stationary and all others execute
out--of--phase oscillation, under fixed or periodic boundary
conditions. All these are in agreement with detailed analytical
results obtained for families of SPOs of the same FPU system under
periodic boundary conditions (see e.g. \cite{Poggi}).

Then, in Sec.~\ref{sec_3}, we vary the values of $E$ and $N$ and
study the behavior of the Lyapunov exponents of the OPMs of the FPU
and BEC Hamiltonians. We find that, as the eigenvalues of the
monodromy matrix exit the unit circle on the real axis, at energies
$0<E_c\equiv E_1<E_2<\ldots$, for fixed $N$, all positive Lyapunov
exponents $L_i,\ i=1,\ldots,N-1$, increase following two distinct
power laws, $L_i\propto E^{B_k}$, $B_k>0,\ k=1,2$, with the $B_k$'s
as reported in the literature, \cite{Rechester,Benettin_1984} and
\cite{Livi}. Furthermore, as the energy $E$ grows at fixed $N$, the
real eigenvalues of the OPM orbit of FPU continue to move away from
$-1$, unlike the OPM of the BEC Hamiltonian, where for very large
$E$ all these eigenvalues tend to return to $+1$. Interestingly
enough, the IPM of the BEC Hamiltonian remains stable for all the
energies and number of dof we studied!

In Sec.~\ref{sec_4}, we turn to the question of the ``size'' of
islands around stable SPOs and use the Smaller Alignment Index
(SALI), introduced in earlier papers
\cite{Skokos_2001,Skokos_1_2003,Skokos_2_2003,Skokos_2004} to
distinguish between regular and chaotic trajectories in our two
Hamiltonians. First, we verify again in these multi--degree of
freedom systems the validity of the SALI dependence on the two
largest Lyapunov exponents $L_1$ and $L_2$ in the case of chaotic
motion, to which it owes its effectiveness and predictive power.
Then, computing the SALI, at points further and further away from
stable SPOs, we determine approximately the ``magnitude'' of these
islands and find that it vanishes (as expected) at the points where
the corresponding OPMs destabilize. In fact, for the OPM of the FPU
system the size of the islands decreases monotonically before
destabilization while for the BEC orbit the opposite happens! Even
more remarkably, for the IPM of the BEC Hamiltonian
(\ref{BEC_Hamiltonian}), not only does the ``size'' of the island
not vanish, it even grows with increasing energy and remains of
considerable magnitude for all the values of $E$ and $N$ we
considered.

Finally, in Sec.~\ref{sec_5}, using as initial conditions the
unstable SPOs, we compute the Lyapunov spectra of the FPU and BEC
systems in the so--called thermodynamic limit, i.e.~as the energy
$E$ and the number of dof $N$ grow indefinitely, with energy density
$\frac{E}{N}$ fixed. First, we find that Lyapunov exponents, fall on
smooth curves of the form $L_{i}\approx L_1e^{-\alpha i/N}$, for
both systems. Then, computing the Kolmogorov--Sinai entropy
$h_{KS}$, as the sum of the positive Lyapunov exponents
\cite{Pesin,Hilborn}, in the thermodynamic limit, we find, for both
Hamiltonians, that $h_{KS}$ is an extensive quantity as it grows
linearly with $N$, demonstrating that in their chaotic regions the
FPU and BEC Hamiltonians behave as ergodic systems of statistical
mechanics.

\section{Simple Periodic Orbits (SPOs) and Local Stability
Analysis}\label{sec_2}
\subsection{The FPU model}\label{sub_sec_2.1}
\subsubsection{Analytical expressions of the OHS mode}\label{sub_sub_sec_2.1.1}
We consider a $1$D lattice of $N$ particles with nearest neighbor
interactions given by the Hamiltonian \cite{Fermi_1955}
\begin{equation}\label{FPU_Hamiltonian_2}
H=\frac{1}{2}\sum_{j=1}^{N}\dot{x}_{j}^{2}+\sum_{j=0}^{N}\biggl(\frac{1}{2}(x_{j+1}-x_{j})^2+\frac{1}{4}\beta(x_{j+1}-x_{j})^4\biggr)
\end{equation}
where $x_{j}$ is the displacement of the $j$th particle from its
equilibrium position, $\dot{x}_{j}$ is the corresponding
canonically conjugate momentum of $x_{j}$ and $\beta$ is a
positive real constant.

Imposing fixed boundary conditions
\begin{equation}\label{FPU_fixed_boundary_conditions_OHS}
x_{0}(t)=x_{N+1}(t)=0,\;\forall t
\end{equation}
one finds a simple periodic orbit first studied by \cite{Ooyama},
taking $N$ odd, which we shall call the OHS mode (using Ooyama's,
Hirooka's and \Saito's initials)
\begin{equation}\label{FPU_non_lin_mode_fixed_boundary_conditions_OHS}
\hat{x}_{2j}(t)=0,\;\hat{x}_{2j-1}(t)=-\hat{x}_{2j+1}(t)\equiv\hat{x}(t),\;j=1,\ldots,\frac{N-1}{2}.
\end{equation}
Here, we shall examine analytically the stability properties of this
mode and determine the energy range $0\leq E\leq E_{c}(N)$ over
which it is linearly stable.

The equations of motion associated with Hamiltonian
(\ref{FPU_Hamiltonian_2}) are
\begin{equation}\label{FPU_eq_motion_OHS}
\ddot{x}_{j}(t)=x_{j+1}-2x_{j}+x_{j-1}+\beta\Bigl((x_{j+1}-x_{j})^3-(x_{j}-x_{j-1})^3\Bigr),\;j=1,\ldots,N
\end{equation}
whence, using the boundary condition
(\ref{FPU_fixed_boundary_conditions_OHS}) and the expressions
(\ref{FPU_non_lin_mode_fixed_boundary_conditions_OHS}) for every
$j=1,3,\ldots,N-2,N$, we arrive at a single equation
\begin{equation}\label{FPU_single_equation_OHS}
\ddot{\hat{x}}(t)=-2\hat{x}(t)-2\beta\hat{x}^{3}(t)
\end{equation}
describing the anharmonic oscillations of all odd particles of the
initial lattice. The solution of (\ref{FPU_single_equation_OHS}) is,
of course, well--known in terms of Jacobi elliptic functions
\cite{Abramowitz} and can be written as
\begin{equation}\label{sol_FPU_single_equation_OHS}
\hat{x}(t)=\mathcal{C}\cn(\lambda t,{\kappa}^{2})
\end{equation}
where
\begin{equation}\label{FPU_C_and_lambda_OHS}
\mathcal{C}^{2}=\frac{2{\kappa}^{2}}{\beta(1-2{\kappa}^{2})},\
\lambda^{2}= \frac{2}{1-2{\kappa}^{2}}
\end{equation}
and ${\kappa}^{2}$ is the modulus of the $\cn$ elliptic function.
The energy per particle of this SPO is then given by
\begin{equation}\label{FPU_energy_per_perticle_OHS}
\frac{E}{N+1}=\frac{1}{4}\mathcal{C}^{2}(2+\mathcal{C}^{2}\beta)=\frac{{\kappa}^{2}-{\kappa}^{4}}{(1-2{\kappa}^{2})^{2}\beta}.
\end{equation}

\subsubsection{Stability analysis of the OHS mode}\label{sub_sub_sec_2.1.2}
Setting $x_{j}=\hat{x}_{j}+y_{j}$ in (\ref{FPU_eq_motion_OHS}) and
keeping up to linear terms in $y_{j}$ we get the corresponding
variational equations for the OHS mode
(\ref{FPU_non_lin_mode_fixed_boundary_conditions_OHS})
\begin{equation}\label{FPU_variational_equations_OHS}
\ddot{y}_{j}=(1+3\beta
\hat{x}^{2})(y_{j-1}-2y_{j}+y_{j+1}),\;j=1,\ldots,N
\end{equation}
where $y_{0}=y_{N+1}=0$.

Using the standard method of diagonalization of linear algebra, we
can separate these variational equations to $N$ uncoupled
independent \Lame equations [Abramowitz \& Stegun, 1965]
\begin{equation}\label{FPU_Lame_equations_OHS}
\ddot{z}_{j}(t)+4(1+3\beta\hat{x}^{2}){\sin}^{2}\biggl(\frac{\pi
j}{2(N+1)}\biggr)z_{j}(t)=0,\;j=1,\ldots,N
\end{equation}
where the $z_{j}$ variations are simple linear combinations of
$y_{j}$'s. Using (\ref{sol_FPU_single_equation_OHS}) and changing
variables to $u=\lambda t$, Eq. (\ref{FPU_Lame_equations_OHS})
takes the form
\begin{equation}\label{FPU_simplified_Lame_equations_OHS}
{z_{j}^{\prime\prime}}(u)+2\bigl(1+4{\kappa}^{2}-6{\kappa}^{2}{\sn}^{2}(u,{\kappa}^{2})\bigr){\sin}^{2}\biggl(\frac{\pi
j}{2(N+1)}\biggr)z_{j}(u)=0,\;j=1,\ldots,N
\end{equation}
where we have used the identity
${\cn}^{2}(u,{\kappa}^{2})=1-{\sn}^{2}(u,{\kappa}^{2})$ and primes
denote differentiation with respect to $u$.

Equation (\ref{FPU_simplified_Lame_equations_OHS}) is an example
of Hill's equation [Copson, $1935$; Magnus \& Winkler, $1966$]
\begin{equation}\label{FPU_Floquet_equation_sample_OHS}
{z^{\prime\prime}}(u)+Q(u)z(u)=0
\end{equation}
where $Q(u)$ is a $T$--periodic function  ($Q(u)=Q(u+T)$) with
$T=2\mathcal{K}$ and $\mathcal{K}\equiv\mathcal{K}(\kappa^{2})$ is
the elliptic integral of the first kind.

According to Floquet theory \cite{Magnus} the solutions of Eq.
(\ref{FPU_Floquet_equation_sample_OHS}) are bounded (or unbounded)
depending on whether the Floquet exponent $\alpha$, given by
\begin{equation}\label{FPU_Floquet_relation_OHS}
\cos\Bigl(2\alpha
\mathcal{K}(\kappa^{2})\Bigr)=1-2{\sin}^{2}\Bigl(\mathcal{K}(\kappa^{2})\sqrt{a_{0}}\Bigr)\det\Bigl(\mathbf{D}(0)\Bigr)
\end{equation}
is real (or imaginary). The matrix $\mathbf{D}(\alpha)$ is called
Hill's matrix and in our case its entries are given by
\begin{equation}\label{FPU_Hill's_matrix_entries_OHS}
[\mathbf{D}(\alpha)]_{n,m}\equiv\frac{a_{n-m}}{a_{0}-\Bigl(\alpha+\frac{n\pi}{\mathcal{K}(\kappa^{2})}\Bigr)^{2}}+\delta_{n,m}
\end{equation}
where $\delta_{n,m}=\begin{cases} 1, & n=m\cr 0,& n\neq m
\cr\end{cases}$, is the Kronecker delta with $n$, $m\in\mathbb{Z}$
and the $a_{n}$'s are the coefficients of the \Four series
expansion of $Q(u)$,
\begin{equation}\label{FPU_series_expansion_Q(u)_OHS}
Q(u)=\sum_{n=-\infty}^{\infty}a_{n}e^{\frac{i n\pi
u}{\mathcal{K}(\kappa^{2})}}.
\end{equation}
Thus, Eq. (\ref{FPU_Floquet_relation_OHS}) gives a stability
criterion for the OHS mode
(\ref{FPU_non_lin_mode_fixed_boundary_conditions_OHS}), by the
condition
\begin{equation}\label{FPU_stability_criterion_OHS}
\Bigl|1-2{\sin}^{2}\Bigl(\mathcal{K}(\kappa^{2})\sqrt{a_{0}}\Bigr)\det\Bigl(\mathbf{D}(0)\Bigr)\Bigr|=\begin{cases}<1,&\mathrm{stable\;
mode}\cr>1,&\mathrm{unstable\;mode}\end{cases}.
\end{equation}
In this case, the \Four coefficients of Hill's matrix
$\mathbf{D}(0)$ are given by the relations [Copson, $1935$]
\begin{eqnarray}\label{FPU_Four_coefficients_OHS}
a_{0}&=&2\Biggr(-5+4\kappa^{2}+6\frac{\mathcal{E}(\kappa^{2})}{\mathcal{K}(\kappa^{2})}\Biggl){\sin}^{2}\Biggl(\frac{\pi
j }{2(N+1)}\Biggl)\\
a_{n}&=&2\frac{6n\pi^{2}q^{n}}{\mathcal{K}^{2}(\kappa^{2})(1-q^{2n})}{\sin}^{2}\Biggl(\frac{\pi
j }{2(N+1)}\Biggl),\;n\neq 0
\end{eqnarray}
where $q\equiv e^{-\pi\frac{\mathcal{K}^{\prime}}{\mathcal{K}}}$,
$\mathcal{K}\equiv\mathcal{K}(\kappa^{2})$ and
$\mathcal{E}\equiv\mathcal{E}(\kappa^{2})$ are the elliptic
integrals of the first and second kind respectively and
$\mathcal{K}^{\prime}\equiv\mathcal{K}(\kappa^{\prime 2})$ with
$\kappa^{\prime 2}\equiv1-\kappa^{2}$.

In the evaluation of $\mathbf{D}(0)$ in Eq.
(\ref{FPU_stability_criterion_OHS}) we have used $121$ terms in
the \Four series expansion of $\sn^{2}(u,\kappa^{2})$ (that is,
$121\times121$ Hill's determinants of $\mathbf{D}(0)$). Thus, we
determine with accuracy $10^{-8}$ the $\kappa^2\equiv
\kappa^{2}_{j}$ values at which the Floquet exponent $\alpha$ in
(\ref{FPU_Floquet_relation_OHS}) becomes zero and the $z_{j}(u)$
in (\ref{FPU_simplified_Lame_equations_OHS}) become unbounded. We
thus find that the first variation $z_{j}(u)$ to become unbounded
as $\kappa^{2}$ increases is $j=\frac{N-1}{2}$ and the energy
values $E_c$ at which this happens (see Eq.
(\ref{FPU_energy_per_perticle_OHS})) are listed in Table
\ref{table_FPU_OHS} for $\beta=1.04$. The $z_{j}(u)$ variation
with $j=\frac{N+1}{2}$ has Floquet exponent $\alpha$ equal to zero
for every $\kappa^{2}\in[0,\frac{1}{2}]$, that is
$z_{\frac{N+1}{2}}(u)$ corresponds to variations along the orbit.
\begin{table}[ht]
\begin{center}
\begin{tabular}{|c|c|c|c|c|c|c|}\hline
$N$ & $5$ & $7$ & $9$ & $11$ & $13$ & $15$\\
\hline $E_c$ &6.4932 &3.0087 &2.2078&1.8596 &1.6669&1.5452\\
\hline
\end{tabular}
\end{center}
\caption{The energies $E_c$, for $\beta=1.04$, at which the
$z_{\frac{N-1}{2}}(u)$ in
(\ref{FPU_simplified_Lame_equations_OHS}) becomes unbounded for
some odd values of $N$.}\label{table_FPU_OHS}
\end{table}

\begin{figure}[ht]
\begin{center}
\includegraphics{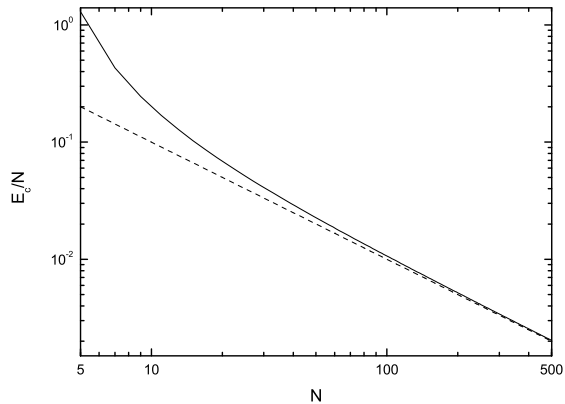} \vspace{4.7cm}
\end{center}
\caption{The solid curve corresponds to the energy per particle
$\frac{E_{c}}{N}$, for $\beta=1.04$, of the first destabilization of
the OHS nonlinear mode
(\ref{FPU_non_lin_mode_fixed_boundary_conditions_OHS}) of the FPU
system (\ref{FPU_Hamiltonian_2}) obtained by the numerical
evaluation of the Hill's determinant in
(\ref{FPU_stability_criterion_OHS}), while the dashed line
corresponds to the function $\propto\frac{1}{N}$. Note that both
axes are logarithmic.}\label{fig_1}
\end{figure}

Next we vary $N$ and calculate the destabilization energy per
particle $\frac{E_{c}}{N}$ for $\beta=1.04$ at which the OHS
nonlinear mode
(\ref{FPU_non_lin_mode_fixed_boundary_conditions_OHS}) becomes
unstable. Plotting the results in Fig. \ref{fig_1}, we see that
$\frac{E_{c}}{N}$ decreases for large $N$ with a simple power--law,
as $1/N$.

\subsubsection{Analytical study of an OPM solution}\label{2.1.3}
We now turn to the properties of another SPO of the FPU Hamiltonian
(\ref{FPU_Hamiltonian_2}), studied in \cite{Budinsky,Poggi} and
[Cafarella {\it et al.}, 2003]. In particular, imposing the periodic
boundary conditions
\begin{equation}\label{FPU_periodic_boundary_conditions_OPM}
x_{N+1}(t)=x_{1}(t),\;\forall t
\end{equation}
with $N$ even, we analyze the stability properties of the
out--of--phase mode (OPM) defined by
\begin{equation}\label{FPU_non_lin_mode_periodic_boundary_conditions_OPM}
\hat{x}_{j}(t)=-\hat{x}_{j+1}(t)\equiv\hat{x}(t),\;j=1,\ldots,N.
\end{equation}
In this case the equations of motion (\ref{FPU_eq_motion_OHS})
reduce also to a single differential equation
\begin{equation}\label{FPU_single_equation_OPM}
\ddot{\hat{x}}(t)=-4\hat{x}(t)-16\beta\hat{x}^{3}(t)
\end{equation}
describing the anharmonic oscillations of all particles of the
initial lattice. The solution of Eq.
(\ref{FPU_single_equation_OPM}) can again be written as an
elliptic $\cn$--function
\begin{equation}\label{sol_FPU_single_equation_OPM}
\hat{x}(t)=\mathcal{C}\cn(\lambda t,{\kappa}^{2})
\end{equation}
with
\begin{equation}\label{FPU_C_and_lambda_OPM}
\mathcal{C}^{2}=\frac{{\kappa}^{2}}{2\beta(1-2{\kappa}^{2})},\ \ \
\lambda^{2}= \frac{4}{1-2{\kappa}^{2}}.
\end{equation}
The energy per particle of the nonlinear OPM
(\ref{FPU_non_lin_mode_periodic_boundary_conditions_OPM}) is given
by
\begin{equation}\label{FPU_energy_per_perticle_OPM}
\frac{E}{N}=2\mathcal{C}^{2}(1+2\mathcal{C}^{2}\beta)=\frac{{\kappa}^{2}-{\kappa}^{4}}{(1-2{\kappa}^{2})^{2}\beta}
\end{equation}
in this case.

We study the linear stability of the OPM
(\ref{FPU_non_lin_mode_periodic_boundary_conditions_OPM}) following
a similar analysis to the one performed in the case of the OHS mode
of Sec.~\ref{sub_sub_sec_2.1.2}. In this case, the corresponding
variational equations have the form
\begin{equation}\label{FPU_variational_equations_OPM}
\ddot{y}_{j}=(1+12\beta
\hat{x}^{2})(y_{j-1}-2y_{j}+y_{j+1}),\;j=1,\ldots,N
\end{equation}
where $y_{1}=y_{N+1}$. After the appropriate diagonalization, the
above equations are transformed to a set of  $N$ uncoupled
independent \Lame equations, which take the form
\begin{equation}\label{FPU_simplified_Lame_equations_OPM}
{z}_{j}^{\prime\prime}(u)+\bigl(1+4{\kappa}^{2}-6{\kappa}^{2}{\sn}^{2}(u,{\kappa}^{2})\bigr){\sin}^{2}\biggl(\frac{\pi
j}{N}\biggr)z_{j}(u)=0,\;j=1,\ldots,N
\end{equation}
after changing to the new time variable $u=\lambda t$. Primes
denotes again differentiation with respect to $u$.

As in Sec.~\ref{sub_sub_sec_2.1.2}, Eq.
(\ref{FPU_stability_criterion_OHS}) gives a stability criterion for
the nonlinear mode
(\ref{FPU_non_lin_mode_periodic_boundary_conditions_OPM}) with
\begin{eqnarray}\label{FPU_Four_coefficients_OPM}
a_{0}&=&\Biggr(-5+4\kappa^{2}+6\frac{\mathcal{E}(\kappa^{2})}{\mathcal{K}(\kappa^{2})}\Biggl){\sin}^{2}\Biggl(\frac{\pi
j }{N}\Biggl),\\
a_{n}&=&\frac{6n\pi^{2}q^{n}}{\mathcal{K}^{2}(\kappa^{2})(1-q^{2n})}{\sin}^{2}\Biggl(\frac{\pi
j }{N}\Biggl),\;n\neq 0.
\end{eqnarray}

Proceeding in the same way as with the OHS mode, we find that the
first variation $z_{j}(u)$ in Eq.
(\ref{FPU_simplified_Lame_equations_OPM}) that becomes unbounded
(for $\beta=1$ and $N$ even) is $j=\frac{N}{2}-1$, in accordance
with \cite{Budinsky}, \cite{Poggi}, \cite{Cafarella} and that this
occurs at the energy values $E_c$, listed in Table
\ref{table_FPU_OPM} for $\beta=1$. The $z_{j}(u)$ variation with
$j=\frac{N}{2}$ corresponds to variations along the orbit.

\begin{table}[ht]
\begin{center}
\begin{tabular}{|c|c|c|c|c|c|c|}\hline
$N$ & $4$ & $6$ & $8$ & $10$ & $12$ & $14$\\
\hline $E_c$ &4.4953 &0.9069 &0.5314 &0.3843 &0.3051 &0.2532\\
\hline
\end{tabular}
\end{center}
\caption{The energies $E_c$, for $\beta=1$, at which the
$z_{\frac{N}{2}-1}(u)$ in
(\ref{FPU_simplified_Lame_equations_OPM}) becomes unbounded for
the first time for some even values of $N$.}\label{table_FPU_OPM}
\end{table}

\begin{figure}[ht]
\begin{center}
\includegraphics{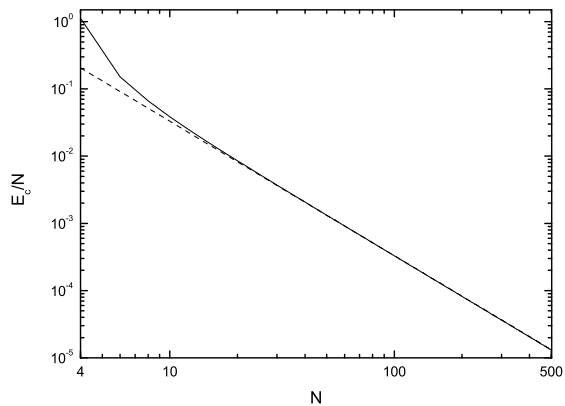} \vspace{4.7cm}
\end{center}
\caption{The solid curve corresponds to the energy per particle
$\frac{E_{c}}{N}$, for $\beta=1$, of the first destabilization of
the nonlinear OPM
(\ref{FPU_non_lin_mode_periodic_boundary_conditions_OPM}) of the FPU
system (\ref{FPU_Hamiltonian_2}) obtained by the numerical
evaluation of the Hill's determinant, while the dashed line
corresponds to the function $\propto\frac{1}{N^2}$. Note that both
axes are logarithmic.}\label{fig_2}
\end{figure}

Taking now many values of N (even) and computing the energy per
particle $\frac{E_{c}}{N}$ for $\beta=1$ at which the OPM
(\ref{FPU_non_lin_mode_periodic_boundary_conditions_OPM}) first
becomes unstable, we plot the results in Fig. \ref{fig_2} and find
that it also decreases following a power--law of the form
$\propto1/N^2$.

\subsection{The BEC model}\label{sub_sec_2.2}
\subsubsection{Analytical expressions for
SPOs}\label{sub_sub_sec_2.2.1} The Hamiltonian of the Bose--Einstein
Condensate (BEC) model studied in this paper is given by
\begin{equation}\label{BEC_Hamiltonian_2}
H=\frac{1}{2}\sum_{j=1}^{N}(p_{j}^{2}+q_{j}^{2})+\frac{\gamma}{8}\sum_{j=1}^{N}(p_{j}^{2}+q_{j}^{2})^2-\frac{\epsilon}{2}\sum_{j=1}^{N}(p_{j}p_{j+1}+q_{j}q_{j+1})
\end{equation}
where $\gamma$ and $\epsilon$ are real constants, which we take here
to be $\gamma=\epsilon=1$. The Hamiltonian (\ref{BEC_Hamiltonian_2})
possesses a second integral of the motion given by
\begin{equation}\label{BEC_Hamiltonian_second_integral}
F=\sum_{j=1}^{N}(p_{j}^{2}+q_{j}^{2})
\end{equation}
and therefore chaotic behavior can only occur for $N\geq3$.

Imposing periodic boundary conditions to the BEC Hamiltonian
(\ref{BEC_Hamiltonian_2})
\begin{eqnarray}\label{BEC_fixed_boundary_conditions}
q_{N+1}(t)&=&q_{1}(t)\;\mathrm{and}\nonumber\\
p_{N+1}(t)&=&p_{1}(t),\;\forall t
\end{eqnarray}
we analyze the stability properties of the in--phase--mode (IPM)
\begin{eqnarray}\label{BEC_non_lin_mode_periodic_boundary_conditions_IPM}
q_{j}(t)&\equiv&\hat{q}(t),\nonumber\\
p_{j}(t)&\equiv&\hat{p}(t)\;\forall
j=1,\ldots,N,\;N\in\mathbb{N}\;\mathrm{and}\;N\ge2
\end{eqnarray}
and of the out--of--phase mode (OPM)
\begin{eqnarray}\label{BEC_non_lin_mode_periodic_boundary_conditions_OPM}
q_{j}(t)&=&-q_{j+1}(t)\equiv\hat{q}(t),\nonumber\\
p_{j}(t)&=&-p_{j+1}(t)\equiv\hat{p}(t)\;\forall
j=1,\ldots,N,\;\mathrm{with}\;N\;\mathrm{only\;even}
\end{eqnarray}
and determine the energy range  $0\leq E\leq E_{c}(N)$ over which
these two SPOs are linearly stable.

In both cases, the corresponding equations of motion
\begin{eqnarray}\label{BEC_eq_motion}
\dot{q}_{j}&=&p_{j}+\frac{\gamma}{2}(p_{j}^{2}+q_{j}^{2})p_{j}-\frac{\epsilon}{2}(p_{j-1}+p_{j+1}),\nonumber\\
\dot{p}_{j}&=&-\Biggl(q_{j}+\frac{\gamma}{2}(p_{j}^{2}+q_{j}^{2})q_{j}-\frac{\epsilon}{2}(q_{j-1}+q_{j+1})\Biggr),\;\forall
j=1,\ldots,N
\end{eqnarray}
give for the IPM solution
\begin{equation}
\dot{\hat{q}}=\hat{p}+\frac{\gamma}{2}(\hat{p}^{2}+\hat{q}^{2})\hat{p}-\epsilon\hat{p},\
\ \
\dot{\hat{p}}=-\biggl(\hat{q}+\frac{\gamma}{2}(\hat{p}^{2}+\hat{q}^{2})\hat{q}-\epsilon\hat{q}\biggr)
\end{equation}
and for the OPM
\begin{equation}
\dot{\hat{q}}=\hat{p}+\frac{\gamma}{2}(\hat{p}^{2}+\hat{q}^{2})\hat{p}+\epsilon\hat{p},\
\ \
\dot{\hat{p}}=-\biggl(\hat{q}+\frac{\gamma}{2}(\hat{p}^{2}+\hat{q}^{2})\hat{q}+\epsilon\hat{q}\biggr).
\end{equation}

From Eq. (\ref{BEC_Hamiltonian_second_integral}) we note that the
second integral becomes for both SPOs
\begin{equation}\label{BEC_Hamiltonian_second_integral_intermediate_relation_1}
F=N(\hat{q}^{2}+\hat{p}^{2})
\end{equation}
yielding for the IPM solution
\begin{equation}
\dot{\hat{q}}=\Biggl(1-\epsilon+\frac{\gamma
F}{2N}\Biggr)^2\hat{p},\ \ \
\dot{\hat{p}}=-\Biggl(1-\epsilon+\frac{\gamma
F}{2N}\Biggr)^2\hat{q}
\end{equation}
and for the OPM
\begin{equation}
\dot{\hat{q}}=\Biggl(1+\epsilon+\frac{\gamma
F}{2N}\Biggr)^2\hat{p},\ \ \
\dot{\hat{p}}=-\Biggl(1+\epsilon+\frac{\gamma
F}{2N}\Biggr)^2\hat{q}.
\end{equation}

The above equations imply for both SPOs that their solutions are
simple trigonometric functions
\begin{eqnarray}
\ddot{\hat{q}}(t)&=&-\omega^2\hat{q}(t)\Rightarrow\nonumber\\
\hat{q}(t)&=&C_{1}\cos(\omega t)+C_{2}\sin(\omega t)\Rightarrow\\
\hat{p}(t)&=&-C_{1}\sin(\omega t)+C_{2}\cos(\omega t)\nonumber
\end{eqnarray}
with $\omega=1-\epsilon+\frac{\gamma F}{2N}$ for the IPM and
$\omega=1+\epsilon+\frac{\gamma F}{2N}$ for the OPM with $C_{1}$
and $C_{2}$ real, arbitrary constants, where $F=NA$ and
$A=C_{1}^{2}+C_{2}^{2}$.

The energy per particle for these two orbits is then given by
\begin{equation}\label{energy_per_particle_BEC_Hamiltonian}
\frac{E}{N}=\frac{1-\epsilon}{2}A+\frac{\gamma}{8}A^{2}\;(\mathrm{IPM})\
\ \ \mathrm{and}\ \ \
\frac{E}{N}=\frac{1+\epsilon}{2}A+\frac{\gamma}{8}A^{2}\;\mathrm{(OPM)}.
\end{equation}

Such SPOs have also been studied in the case of the integrable
so--called dimer problem by other authors \cite{Aubry}, who were
interested in comparing the classical with the quantum properties of
the BEC Hamiltonian (\ref{BEC_Hamiltonian_2}).

\subsubsection{Stability analysis of the SPOs}\label{2.2.2}
Setting now
\begin{eqnarray}
q_{j}&=&\hat{q}_{j}+x_{j},\nonumber\\
p_{j}&=&\hat{p}_{j}+y_{j},\;\forall j=1,\ldots,N
\end{eqnarray}
in the equations of motion (\ref{BEC_eq_motion}) and keeping up to
linear terms in $x_{j}$ and $y_{j}$ we get the corresponding
variational equations for both SPOs
\begin{eqnarray}\label{BEC_linearized_equations}
\dot{x}_{j}&=&\biggl(-\frac{\epsilon}{2}\biggr)y_{j-1}+L_{+}y_{j}-\biggl(\frac{\epsilon}{2}\biggr)y_{j+1}+Kx_{j},\nonumber\\
\dot{y}_{j}&=&-\biggl(\biggl(-\frac{\epsilon}{2}\biggr)x_{j-1}+L_{-}x_{j}-\biggl(\frac{\epsilon}{2}\biggr)x_{j+1}+Ky_{j}\biggr),\;\forall
j=1,\ldots,N
\end{eqnarray}
where $x_{0}=x_{N},y_{0}=y_{N}$ and $x_{N+1}=x_{1},y_{N+1}=y_{1}$
(periodic boundary conditions) and
\begin{eqnarray}\label{BEC_linear_coupled_system}
K&=&\gamma\hat{p}_{j}\hat{q}_{j}=\frac{1}{2}\gamma(2C\cos(2\omega
t)+B\sin(2\omega t)),\nonumber\\
L_{+}&=&1+\frac{\gamma}{2}(\hat{p}_{j}^{2}+\hat{q}_{j}^{2})+\gamma\hat{p}_{j}^{2}=1+A\gamma+\frac{1}{2}B\gamma\cos(2\omega
t)-C\gamma\sin(2\omega t),\\
L_{-}&=&1+\frac{\gamma}{2}(\hat{p}_{j}^{2}+\hat{q}_{j}^{2})+\gamma\hat{q}_{j}^{2}=1+A\gamma-\frac{1}{2}B\gamma\cos(2\omega
t)+C\gamma\sin(2\omega t)\nonumber
\end{eqnarray}
where $B=C_{2}^{2}-C_{1}^{2}$ and $C=C_{1}C_{2}$ are real
constants.

Unfortunately, it is not as easy to uncouple this linear system of
differential equations (\ref{BEC_linearized_equations}), as it was
in the FPU case, in order to study analytically the linear stability
of these two SPOs. We can, however, compute with arbitrarily
accuracy for every given $N$ the eigenvalues of the corresponding
monodromy matrix of the IPM and OPM solutions of the BEC Hamiltonian
(\ref{BEC_Hamiltonian_2}).

Thus, in the case of the OPM
(\ref{BEC_non_lin_mode_periodic_boundary_conditions_OPM}) we
computed for some even values of $N$ the energy thresholds $E_c$ at
which this SPO becomes unstable (see Table \ref{table_BEC_OPM}).

\begin{table}[ht]
\begin{center}
\begin{tabular}{|c|c|c|c|c|c|c|}\hline
$N$ & $2$ & $4$ & $6$ & $8$ & $10$ & $12$\\
\hline $E_c$ &5.0000 &4.5000 &3.1875 &2.4289 &1.9554 &1.6346\\
\hline
\end{tabular}
\end{center}
\caption{The energy $E_c$ of the BEC Hamiltonian
(\ref{BEC_Hamiltonian_2}) at which the OPM
(\ref{BEC_non_lin_mode_periodic_boundary_conditions_OPM}) first
becomes unstable for some even values of $N$.}\label{table_BEC_OPM}
\end{table}

\begin{figure}[ht]
\begin{center}
\includegraphics{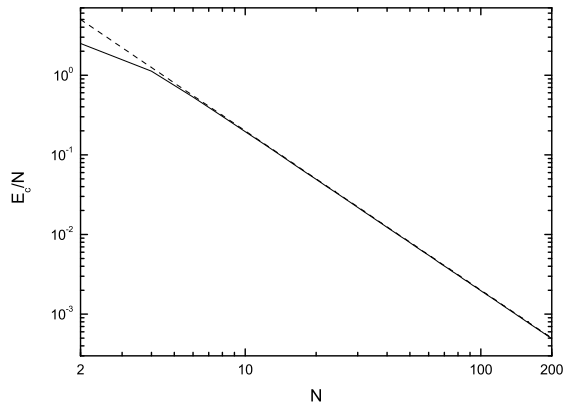} \vspace{4.7cm}
\end{center}
\caption{The solid curve corresponds to the energy per particle
$\frac{E_{c}}{N}$ of the first destabilization of the OPM
(\ref{BEC_non_lin_mode_periodic_boundary_conditions_OPM}) of the BEC
Hamiltonian (\ref{BEC_Hamiltonian_2}) obtained by the numerical
evaluation of the eigenvalues of the monodromy matrix of Eq.
(\ref{BEC_linearized_equations}), while the dashed line corresponds
to the function $\propto\frac{1}{N^{2}}$. Note that both axes are
logarithmic.}\label{fig_3}
\end{figure}

Plotting the results in Fig. \ref{fig_3} we observe again that
$E_c/N$ decreases with $N$ following a power--law $\propto 1/N^{2}$
as in the case of the OPM of the FPU model. On the other hand, the
IPM orbit (\ref{BEC_non_lin_mode_periodic_boundary_conditions_IPM})
was found to remain stable for all the values of $N$ and $E$ we
studied (up to $N=54$ and $E\approx10^5$).

\section{Destabilization of SPOs and Globally Chaotic Dynamics}\label{sec_3}
Let us now study the chaotic behavior in the neighborhood of our
unstable SPOs, starting with the well--known method of the
evaluation of the spectrum of Lyapunov Exponents (LEs) of a
Hamiltonian dynamical system, $L_i,\;i=1,\ldots,2N$ where $L_1\equiv
L_{\mathrm{max}}>L_2>\ldots>L_{2N}$. The LEs measure the rate of
exponential divergence of initially nearby orbits in the phase space
of the dynamical system as time approaches infinity. In Hamiltonian
systems, the LEs come in pairs of opposite sign, so their sum
vanishes, $\sum_{i=1}^{2N}L_i=0$ and two of them are always equal to
zero corresponding to deviations along the orbit under
consideration. If at least one of them (the largest one) $L_1\equiv
L_{\mathrm{max}}>0$, the orbit is chaotic, i.e. almost all nearby
orbits diverge exponentially in time, while if $L_{\mathrm{max}}=0$
the orbit is stable (linear divergence of initially nearby orbits).
Benettin {\it et al.}, [$1980a,\;b$] studied in detail the problem
of the computation of all LEs and proposed an efficient algorithm
for their numerical computation, which we use here.

In particular, $L_i\equiv L_i(\vec{x}(t))$ for a given orbit
$\vec{x}(t)$ is computed as the limit for $t \rightarrow\infty$ of
the quantities
\begin{eqnarray}
K_{t}^i&=&\frac{1}{t}\,\ln\frac{\parallel\vec{w_{i}}(t)\parallel}{\parallel\vec{w_i}(0)\parallel},\\
L_{i}&=&\lim_{t\rightarrow \infty}K_{t}^i\,
\label{Lyap._exps._definition}
\end{eqnarray}
where $\vec{w_i}(0)$ and $\vec{w_i}(t),\;i=1,\ldots,2N$ are
deviation vectors from the given orbit $\vec{x}(t)$, at times
$t=0$ and $t>0$ respectively. The time evolution of $\vec{w_i}$ is
given by solving the so--called variational equations, i.e. the
linearized equations about the orbit. Generally, for almost all
choices of initial deviations $\vec{w_i}(0)$, the limit $t
\rightarrow \infty$ of Eq. (\ref{Lyap._exps._definition}) gives
the same $L_i$.

In practice, of course, since the exponential growth of
$\vec{w_i}(t)$ occurs for short time intervals, one stops the
evolution of $\vec{w_i}(t)$ after some time $T_1$, records the
computed $K_{T_1}^i$, orthogonormalizes the vectors $\vec{w_i}(t)$
and repeats the calculation for the next time interval $T_2$,
etc.~obtaining finally $L_i$ as an average over many $T_j$,
$j=1,2,\ldots,n$ given by
\begin{equation}
L_i=\frac{1}{n} \sum_{j=1}^{n} K_{T_j}^i,\;n\rightarrow\infty.
\end{equation}

\begin{figure}[ht]
\begin{center}
\includegraphics{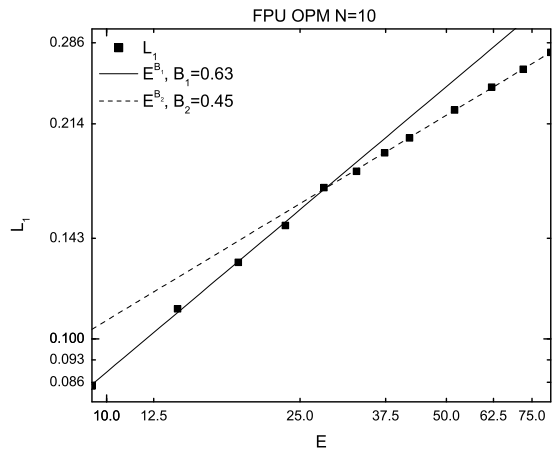} \vspace{4.9cm}
\end{center}
\caption{The two distinct power--law behaviors in the evolution of
the maximum Lyapunov exponent $L_1$ as the energy grows for the OPM
(\ref{FPU_non_lin_mode_periodic_boundary_conditions_OPM}) of the FPU
Hamiltonian (\ref{FPU_Hamiltonian_2}) for $N=10$. A similar picture
is obtained for the $L_2$ and $L_3$ also, with similar exponents and
the transition occurring at the same energy value (see text). Note
that both axes are logarithmic.}\label{fig_4}
\end{figure}

Next, we varied the values of the energy $E$ keeping $N$ fixed and
studied the behavior of the Lyapunov exponents, using as initial
conditions the OPMs
(\ref{FPU_non_lin_mode_periodic_boundary_conditions_OPM}) and
(\ref{BEC_non_lin_mode_periodic_boundary_conditions_OPM}) of the FPU
and BEC Hamiltonians respectively.  First, we find that the values
of the maximum Lyapunov exponent $L_1$ increase by two distinct
power--law behaviors ($L_1\propto E^B,\ B>0$) as is clearly seen in
Fig.~\ref{fig_4} for the OPM of the FPU system. The result for the
$L_1$ for the power--law behavior shown by solid line in
Fig.~\ref{fig_4} is in agreement with the results in
\cite{Rechester} and \cite{Benettin_1984}, where they obtain $B=0.5$
and $B=2/3$ for low dimensional systems and differs slightly from
the one obtained in \cite{Livi} for the higher--dimensional case of
$N=80$. We also find the same power--law behaviors with similar
exponents $B$ for the other positive Lyapunov exponents as well. For
example, for the $L_2$ we obtain $L_2\propto E^{0.62}$ and
$L_2\propto E^{0.48}$ and $L_3\propto E^{0.68}$ and $L_3\propto
E^{0.49}$, with the transition occurring at $E\cong28.21$ for all of
them.

\begin{figure}[ht]
\begin{centering}
\includegraphics{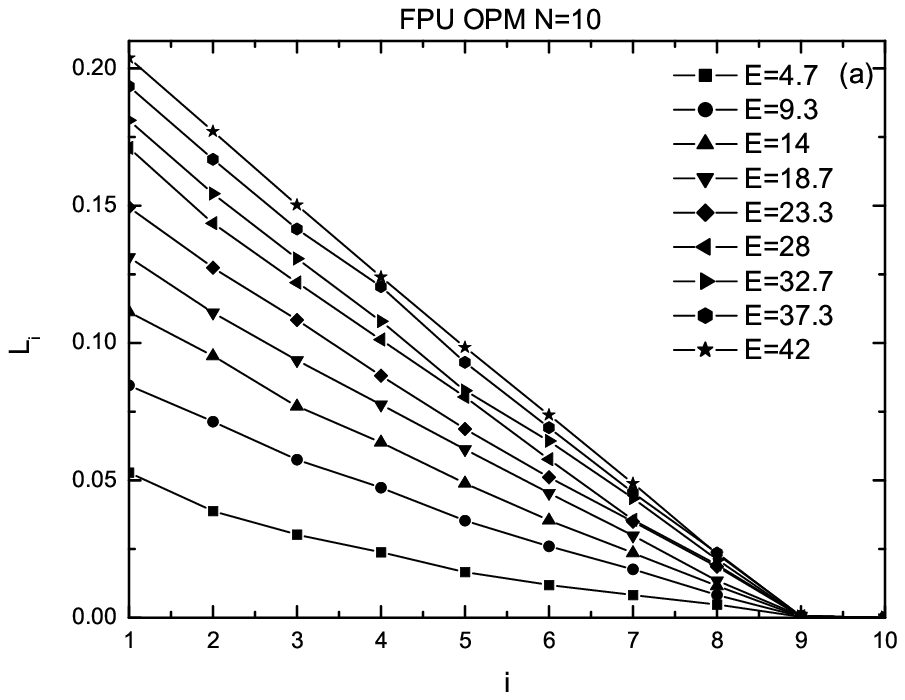} \includegraphics{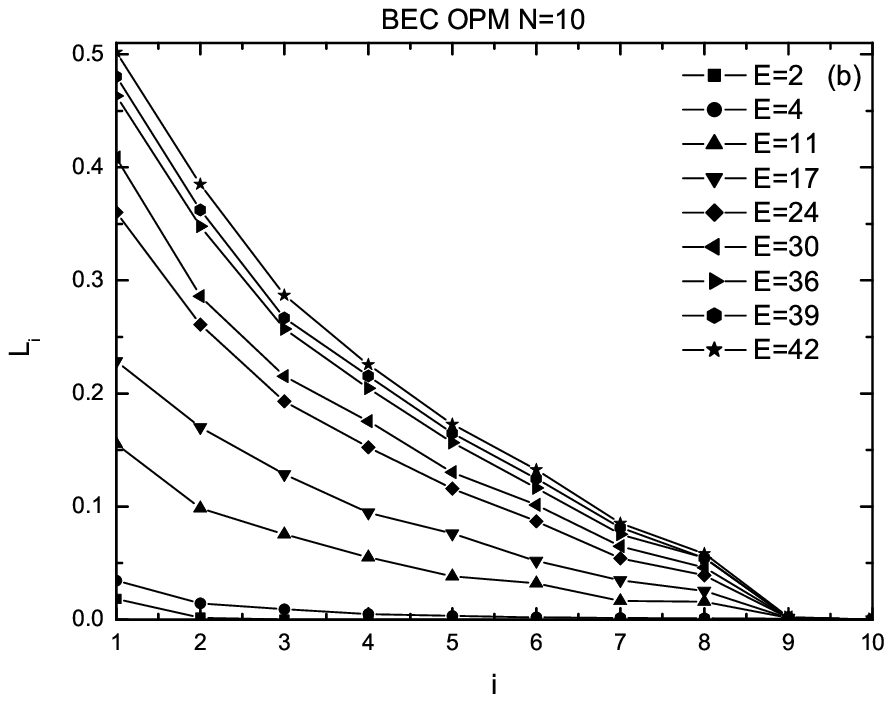} \includegraphics{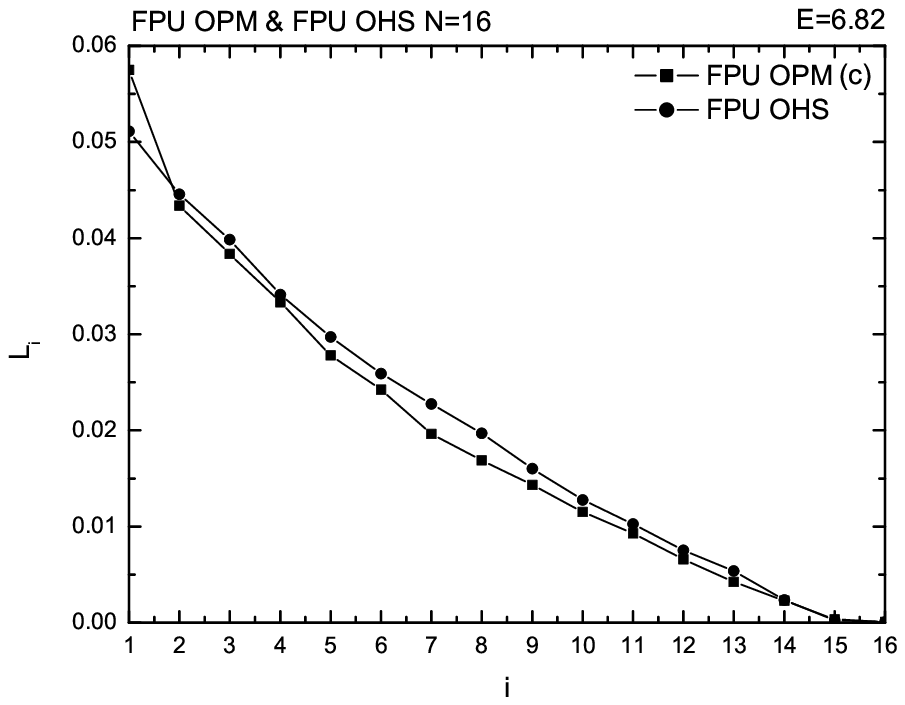} \includegraphics{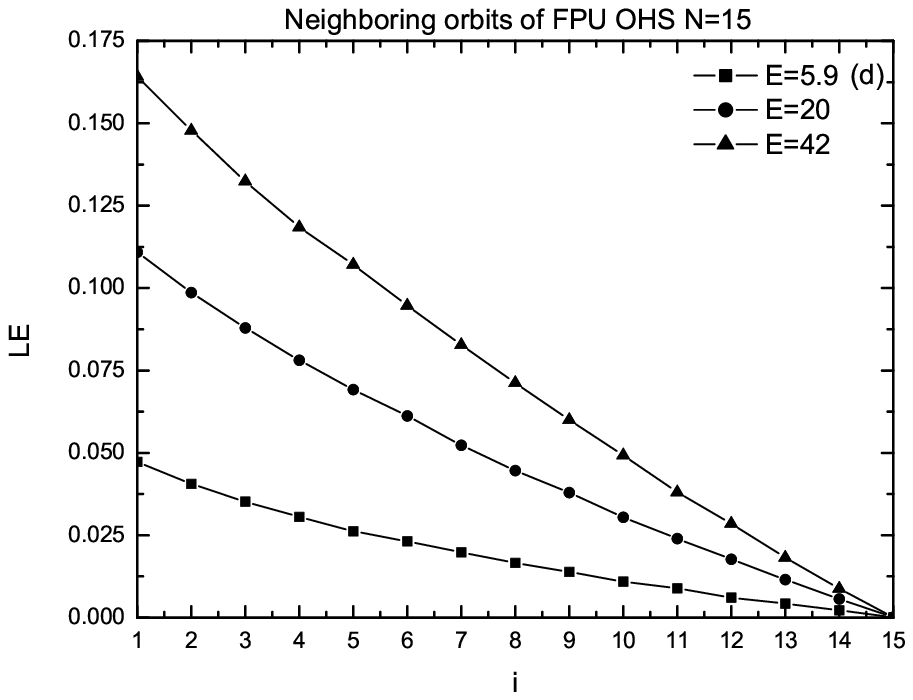} \vspace{8.2cm}
\end{centering}
\caption{(a) The spectrum of the positive Lyapunov exponents for
fixed $N=10$ of the OPM
(\ref{FPU_non_lin_mode_periodic_boundary_conditions_OPM}) of the FPU
Hamiltonian (\ref{FPU_Hamiltonian_2}) as the energy grows. (b) Also
for $N=10$, the OPM
(\ref{BEC_non_lin_mode_periodic_boundary_conditions_OPM}) of the BEC
Hamiltonian (\ref{BEC_Hamiltonian_2}) yields a similar picture as
the energy is increased. (c) The Lyapunov spectrum of the OPM
(\ref{FPU_non_lin_mode_periodic_boundary_conditions_OPM}) of the FPU
Hamiltonian (\ref{FPU_Hamiltonian_2}) for $N=16$ and the OHS mode
(\ref{FPU_non_lin_mode_fixed_boundary_conditions_OHS}) of the same
Hamiltonian and $N$, for periodic boundary conditions practically
coincide at $E=6.82$ where both of them are destabilized. (d) The
Lyapunov spectrum of the FPU OHS mode
(\ref{FPU_non_lin_mode_fixed_boundary_conditions_OHS}) with fixed
boundary conditions for $N=15$ as the energy grows presents as shape
which is qualitatively similar to what was found for the SPOs of
panel (c).} \label{fig_5}
\end{figure}

Turning now to the full Lyapunov spectrum in Fig. \ref{fig_5} we see
that, for fixed $N$, as the energy is increased (and more
eigenvalues of the monodromy matrix exit the unit circle) the
Lyapunov spectrum tends to fall on a smooth curve for the OPM orbits
of FPU and BEC, see Fig. \ref{fig_5}(a), (b), as well as for the OHS
mode with periodic boundary conditions (Fig. \ref{fig_5}(c)).
Observe that in Fig. \ref{fig_5}(c) we have plotted the Lyapunov
spectrum of both the OPM
(\ref{FPU_non_lin_mode_periodic_boundary_conditions_OPM}) of the FPU
Hamiltonian (\ref{FPU_Hamiltonian_2}) and of the OHS mode
(\ref{FPU_non_lin_mode_fixed_boundary_conditions_OHS}) for $N=16$
and periodic boundary conditions at the energy $E=6.82$ where both
of them are destabilized and their distance in phase space is such
that they are far away from each other. We clearly see that the two
Lyapunov spectra are almost identical suggesting that their chaotic
regions are somehow ``connected'', as orbits starting initially in
the vicinity of one of these SPOs visit often in the course of time
the chaotic region of the other one. In Fig. \ref{fig_5}(d) we have
plotted the positive Lyapunov exponents spectra of three neighboring
orbits of the OHS mode for $N=15$ dof in three different energies
and observe that the curves are qualitatively the same as in Fig.
\ref{fig_5}(a) and (c).

\begin{figure}[ht]
\begin{center}
\includegraphics{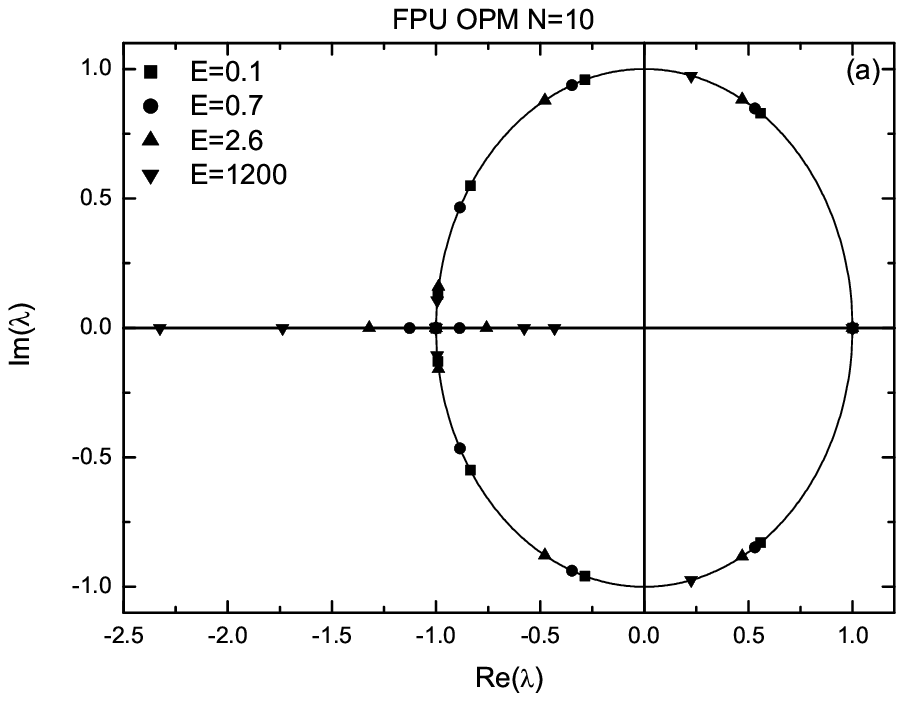} \includegraphics{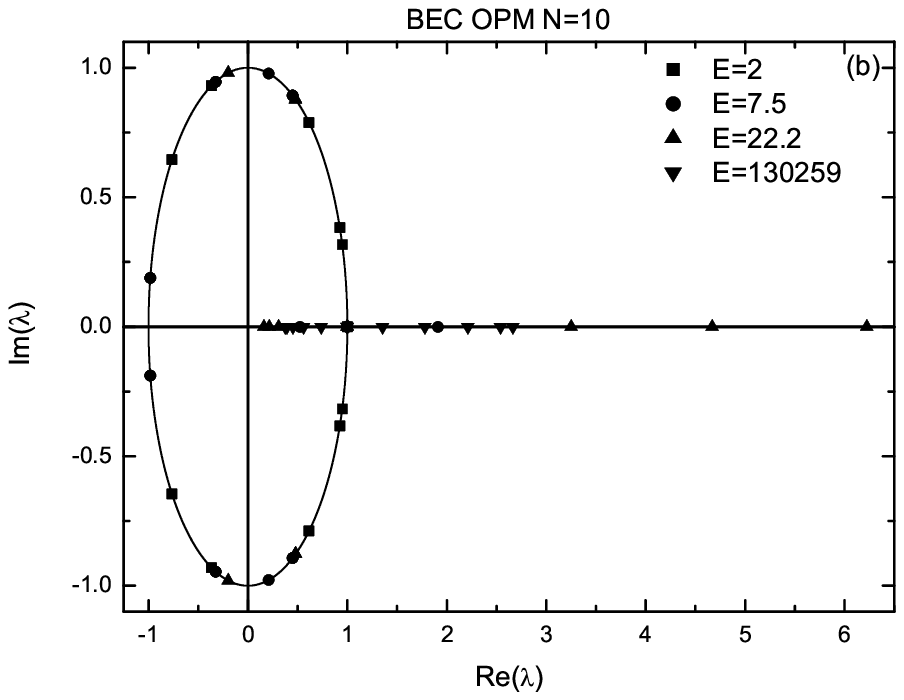} \includegraphics{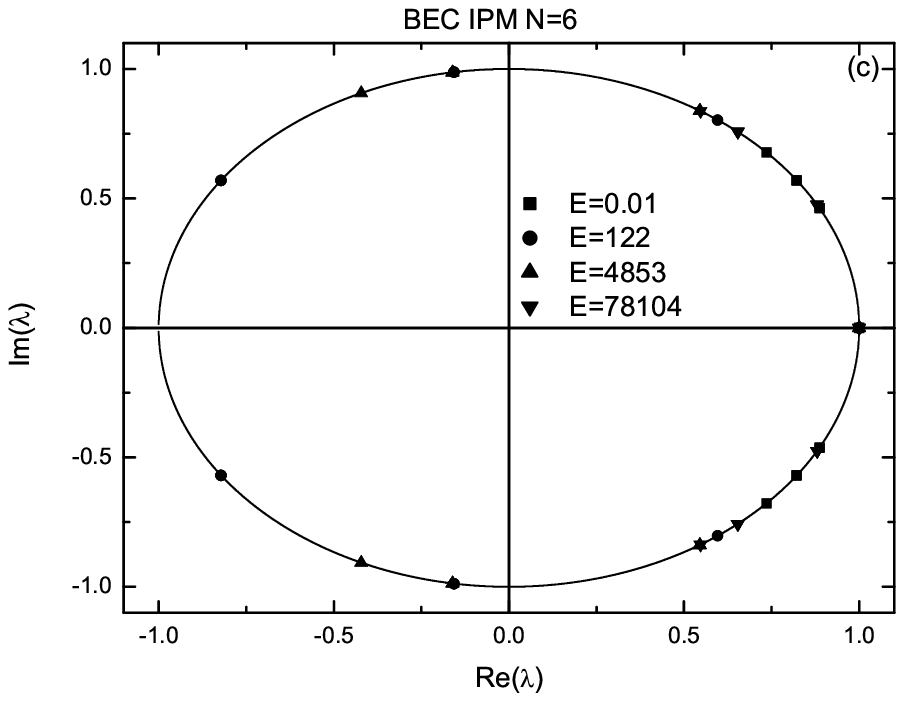} \vspace{8cm}
\end{center}
\caption{The eigenvalues $\lambda_{j},\;j=1,\ldots,2N$ (a) of the
OPM (\ref{FPU_non_lin_mode_periodic_boundary_conditions_OPM}) of the
FPU Hamiltonian (\ref{FPU_Hamiltonian_2}) with $N=10$ dof. (b) The
eigenvalues of the OPM
(\ref{BEC_non_lin_mode_periodic_boundary_conditions_OPM}) of the BEC
Hamiltonian (\ref{BEC_Hamiltonian_2}) with the same number of dof
$N$. (c) The eigenvalues of the IPM
(\ref{BEC_non_lin_mode_periodic_boundary_conditions_IPM}) of the BEC
Hamiltonian (\ref{BEC_Hamiltonian_2}) with $N=6$ dof.} \label{fig_6}
\end{figure}

Finally, in Fig. \ref{fig_6} we have plotted the eigenvalues of the
monodromy matrix of several SPOs and have observed the following:
For the OPM of the FPU Hamiltonian the eigenvalues exit from $-1$
(as the orbit destabilizes via period--doubling) and continue to
move away from the unit circle, as $E$ increases further, see Fig.
\ref{fig_6}(a). By contrast, the eigenvalues of the OPM of the BEC
Hamiltonian exit from $+1$ by a symmetry breaking bifurcation and
for very large $E$ tend to return again to $+1$, see Fig.
\ref{fig_6}(b). This does not represent, however, a return to
globally regular motion around this SPO, as the Lyapunov exponents
in its neighborhood remain far from zero. Finally, in Fig.
\ref{fig_6}(c), we show an example of the fact that the eigenvalues
of the IPM orbit of the BEC system, remain all on the unit circle,
no matter how high the value of the energy is. Here, $N=6$, but a
similar picture occurs for all the other values of $N$ we have
studied up to $N=54$ and $E\approx10^5$.

\section{Using SALI to Estimate the ``Size'' of Islands of Regular Motion}\label{sec_4}
In this section, we  estimate the ``size'' of islands of regular
motion around stable SPOs using the Smaller Alignment Index (SALI)
method \cite{Skokos_2001, Skokos_1_2003, Skokos_2_2003,
Skokos_2004}, to distinguish between regular and chaotic orbits in
the FPU and BEC Hamiltonians. The computation of the SALI has proved
to be a very efficient method in revealing rapidly and with
certainty the regular vs. chaotic nature of orbits, as it exhibits a
completely different behavior for the two cases: It fluctuates
around non--zero values for regular orbits, while it converges
exponentially to zero for chaotic orbits. The behavior of the SALI
for regular motion was studied and explained in detail by
\cite{Skokos_2_2003}, while a more analytical study of the behavior
of the index in the case of chaotic motion can be found in
\cite{Skokos_2004}.

As a first step, let us verify in the case of chaotic orbits of
our $N$--degree of freedom systems, the validity of SALI's
dependence on the two largest Lyapunov exponents $L_1\equiv
L_{\mathrm{max}}$ and $L_2$ proposed and numerically checked for
$N=2$ and $3$, in \cite{Skokos_2004}
\begin{equation}\label{SALI_MLE_relation}
\mathrm{SALI(t)}\propto e^{-(L_{1}-L_{2})t}.
\end{equation}
This expression is very important as it implies that chaotic
behavior can be decided by the exponential decay of this parameter,
rather than the often questionable convergence of Lyapunov exponents
to a positive value.

\begin{figure}[ht]
\begin{centering}
\includegraphics{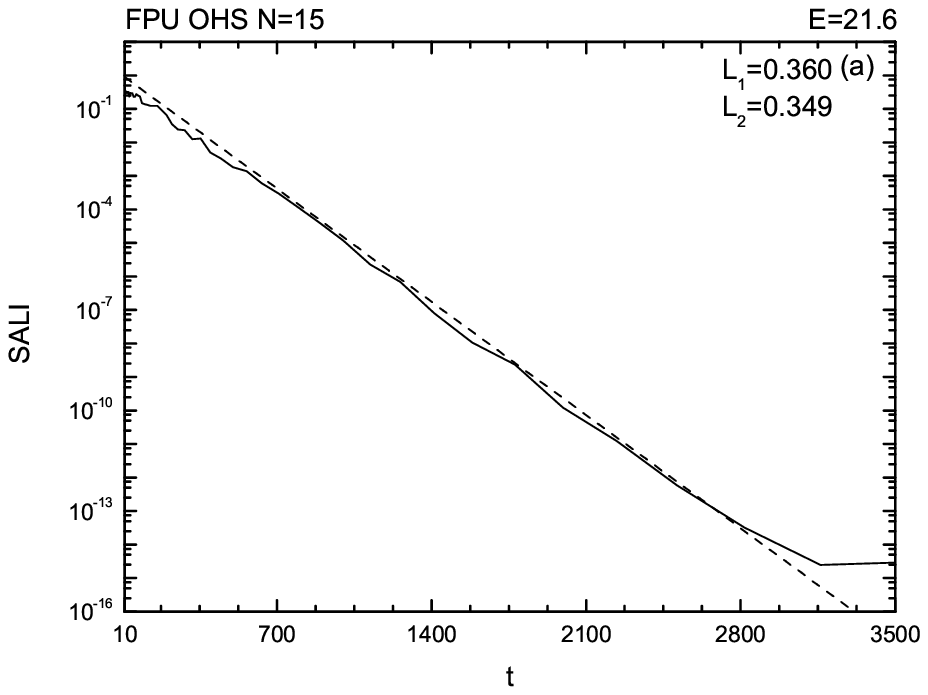} \includegraphics{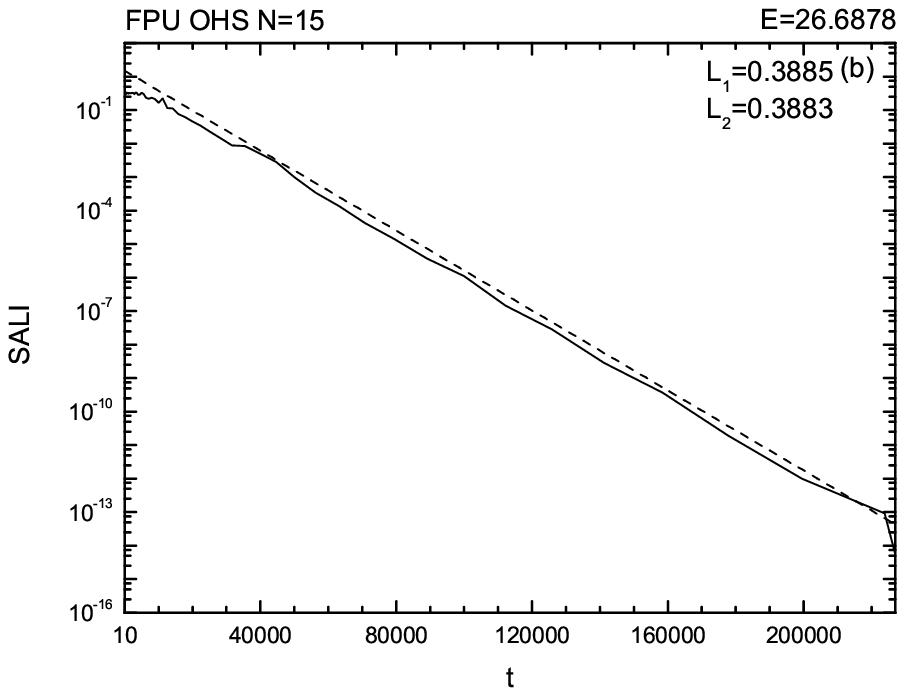} \vspace{4cm}
\end{centering}
\caption{(a) The time evolution of the SALI (solid curve) and of the
Eq.~(\ref{SALI_MLE_relation}) (dashed line) at $E=21.6$ of the OHS
mode (\ref{FPU_non_lin_mode_fixed_boundary_conditions_OHS}) of the
FPU Hamiltonian (\ref{FPU_Hamiltonian_2}) with fixed boundary
conditions. (b) Similar plot to panel (a) but for the larger energy
$E=26.6878$ for which the two largest Lyapunov exponents $L_1$ and
$L_2$ are almost equal while all the other positive ones are very
close to zero. In both panels the agreement between the data (solid
curve) and the derived function of Eq.~(\ref{SALI_MLE_relation})
(dashed line) is remarkably good. Note that the horizontal axes in
both panels are linear.} \label{fig_7}
\end{figure}

To check the validity of (\ref{SALI_MLE_relation}) let us take as an
example the OHS mode
(\ref{FPU_non_lin_mode_fixed_boundary_conditions_OHS}) of the FPU
Hamiltonian (\ref{FPU_Hamiltonian_2}) using fixed boundary
conditions, with $N=15$ dof and $\beta=1.04$, at the energy $E=21.6$
and calculate the Lyapunov exponents, as well as the corresponding
SALI evolution. Plotting SALI as a function of time $t$ (in linear
scale) together with its analytical formula
(\ref{SALI_MLE_relation}) in Fig. \ref{fig_7}(a), we see indeed an
excellent agreement. Increasing further the energy to the value
$E=26.6878$, it is in fact possible to verify expression
(\ref{SALI_MLE_relation}), even in the case where the two largest
Lyapunov exponents are nearly equal, as Fig. \ref{fig_7}(b)
evidently shows!

Exploiting now the different behavior of SALI for regular and
chaotic orbits, we estimate approximately the ``size'' of regions of
regular motion (or, ``islands'' of stability) in phase space, by
computing SALI at points further and further away from a stable
periodic orbit checking whether the orbits are still on a torus
(SALI$\geq10^{-8}$) or have entered a chaotic ``sea''
(SALI$<10^{-8}$) up to the integration time $t=4000$. The initial
conditions are chosen perturbing all the positions of the stable SPO
by the same quantity $dq$ and all the canonically conjugate momenta
by the same $dp$ while keeping always constant the integral $F$,
given by Eq. (\ref{BEC_Hamiltonian_second_integral}), in the case of
the BEC Hamiltonian (\ref{BEC_Hamiltonian_2}) and the energy $E$ in
the case of the FPU Hamiltonian (\ref{FPU_Hamiltonian_2}). In this
way, we are able to estimate the approximate ``magnitude'' of the
islands of stability for the OPM of Hamiltonian
(\ref{FPU_Hamiltonian_2}) and for the IPM and OPM of Hamiltonian
(\ref{BEC_Hamiltonian_2}) varying the energy $E$ and the number of
dof $N$.

\begin{figure}[ht]
\begin{center}
\includegraphics{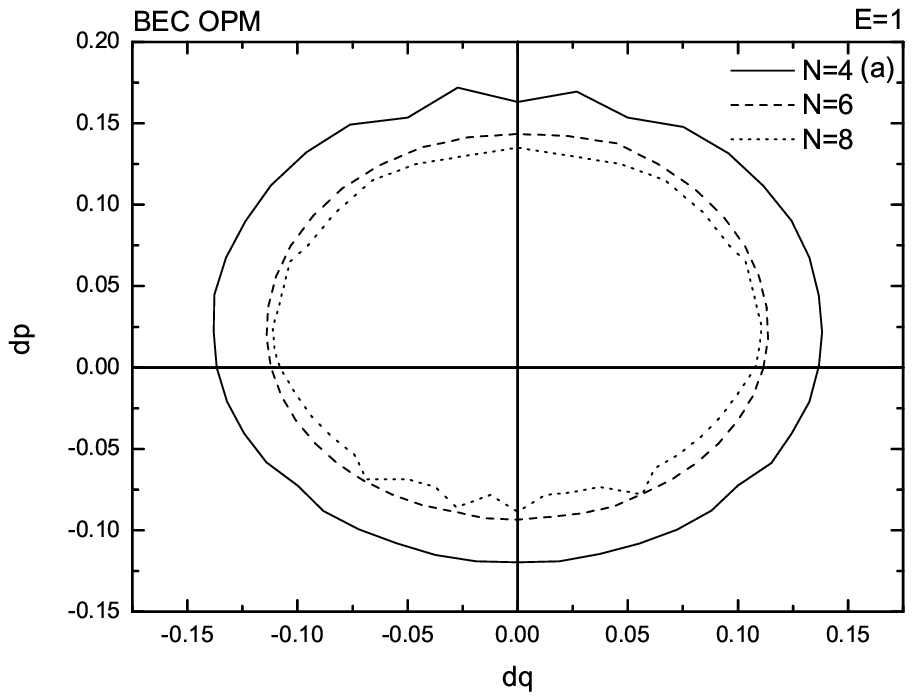} \includegraphics{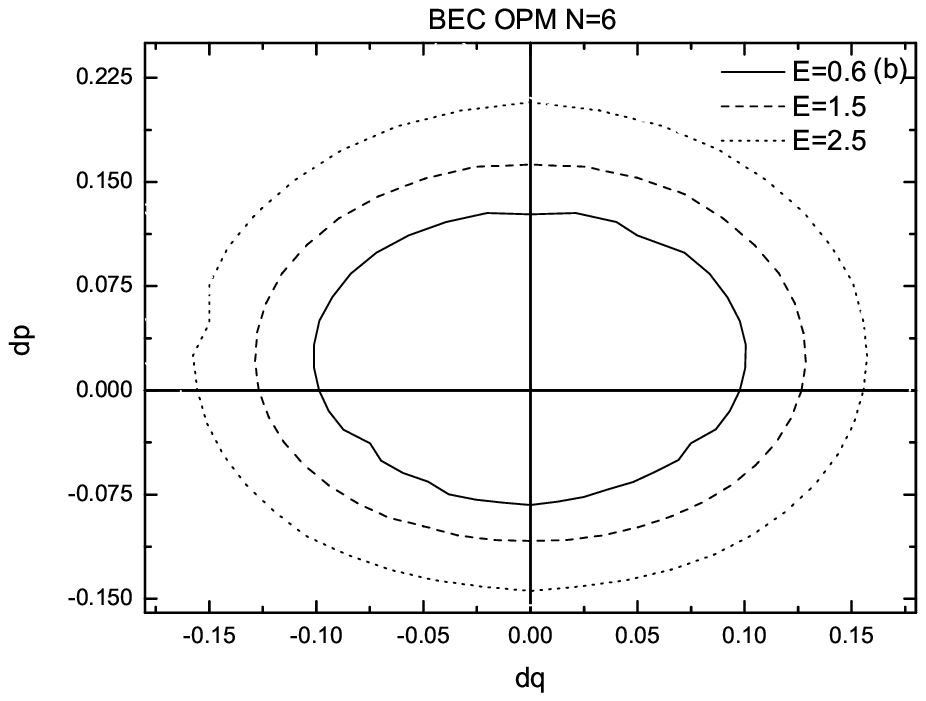} \includegraphics{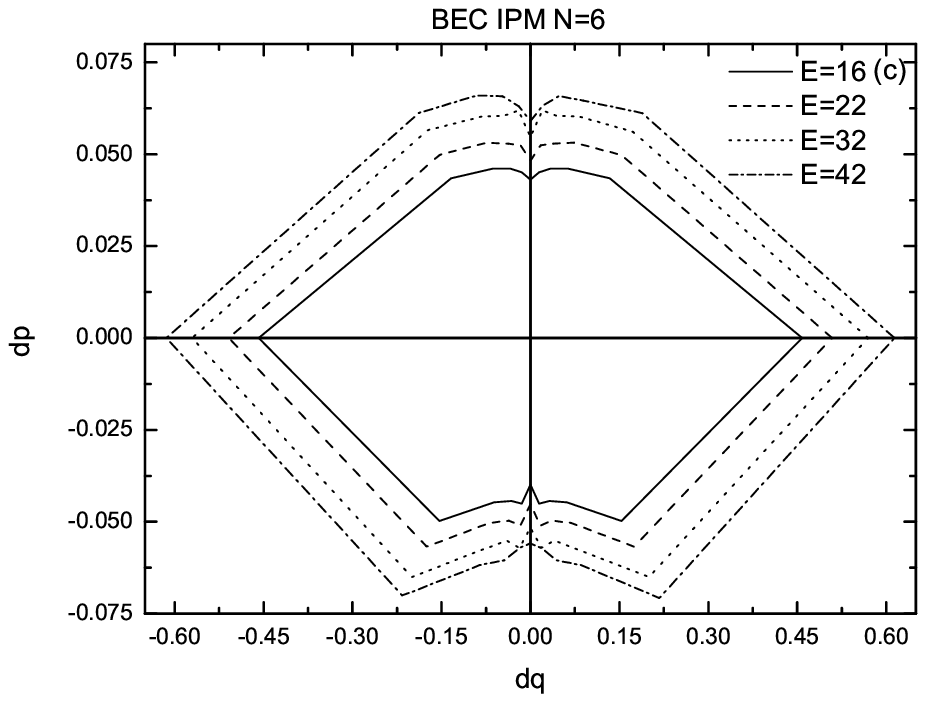} \includegraphics{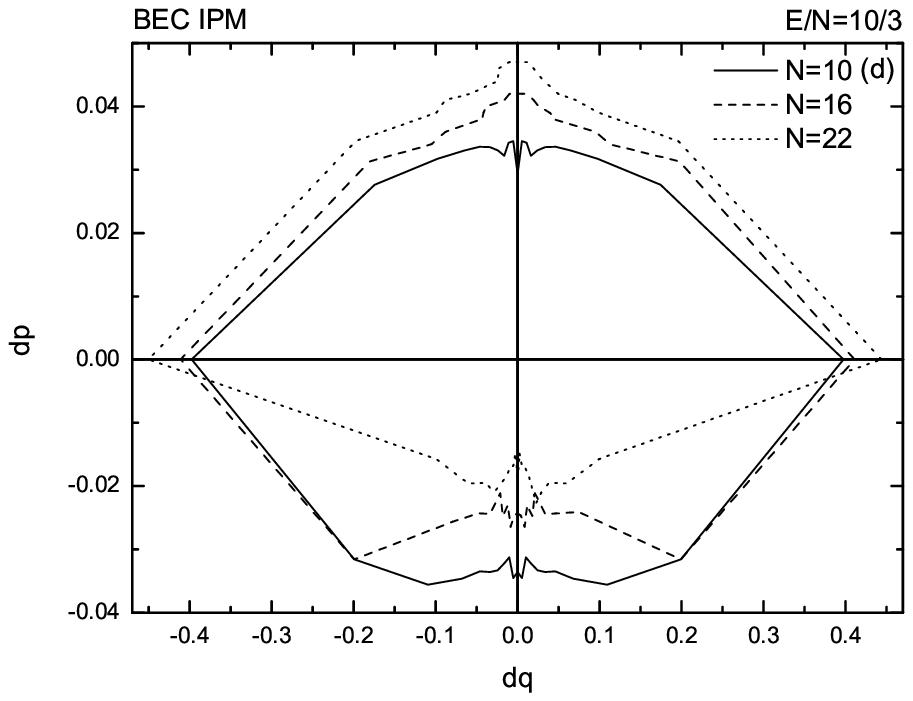} \vspace{8cm}
\end{center}
\caption{(a) ``Size'' of the islands of stability of the OPM
(\ref{BEC_non_lin_mode_periodic_boundary_conditions_OPM}) of the BEC
Hamiltonian (\ref{BEC_Hamiltonian_2}) for $N=4,\ 6$ and $8$ dof and
SPOs constant energy $E=1$ before the first destabilization (see
Table \ref{table_BEC_OPM}). (b) ``Size'' of the islands of stability
of the same Hamiltonian and SPO as in (a) for $N=6$ dof and three
different energies of the SPO before the first destabilization (see
Table \ref{table_BEC_OPM}). (c) ``Size'' of the islands of stability
of the same Hamiltonian as in (a) of the IPM
(\ref{BEC_non_lin_mode_periodic_boundary_conditions_IPM}) for $N=6$
dof and four different energies of the SPO. (d) ``Size'' of the
islands of stability of the same Hamiltonian as in (a) of the IPM
(\ref{BEC_non_lin_mode_periodic_boundary_conditions_IPM}) for
$\frac{E}{N}=\frac{10}{3}$. Here $E$ corresponds to the energy of
the IPM.} \label{fig_8}
\end{figure}

In the case of the OPM solutions of both Hamiltonians, as the number
of dof $N$ increases, for fixed energy $E$, the islands of stability
eventually shrink to zero and the SPOs destabilize. For example,
this is seen in Fig. \ref{fig_8}(a) for the islands around the OPM
(\ref{BEC_non_lin_mode_periodic_boundary_conditions_OPM}) of the BEC
Hamiltonian (\ref{BEC_Hamiltonian_2}). A surprising behavior,
however, is observed for the same SPO, if we keep $N$ fixed and
increase the energy: Instead of diminishing, as expected from the
FPU and other examples, the island of stability actually grows, as
shown in Fig.~\ref{fig_8}(b), for the case of $N=6$ dof. In fact, it
remains of considerable size until the SPO is destabilized for the
first time, through period--doubling bifurcation at
$E\approx3.1875$, whereupon the island ceases to exist!

But what happens to the island of stability around the IPM solution
of the BEC Hamiltonian (\ref{BEC_Hamiltonian_2}), which does not
become unstable for all values of $N$ and $E$ we studied? Does it
shrink to zero at sufficiently large $E$ or $N$? From
Fig.~\ref{fig_8}(c) we see that for a fixed value of $N$, the size
of this island also increases as the energy increases. In fact, this
SPO has large islands about it even if the energy is increased,
keeping the ratio $E/N$ constant (see Fig.~\ref{fig_8}(d)). This was
actually found to be true for considerably larger $E$ and $N$ values
than shown in this figure.

\section{Lyapunov Spectra and the Thermodynamic Limit}\label{sec_5}
\begin{figure}[ht]
\begin{centering}
\includegraphics{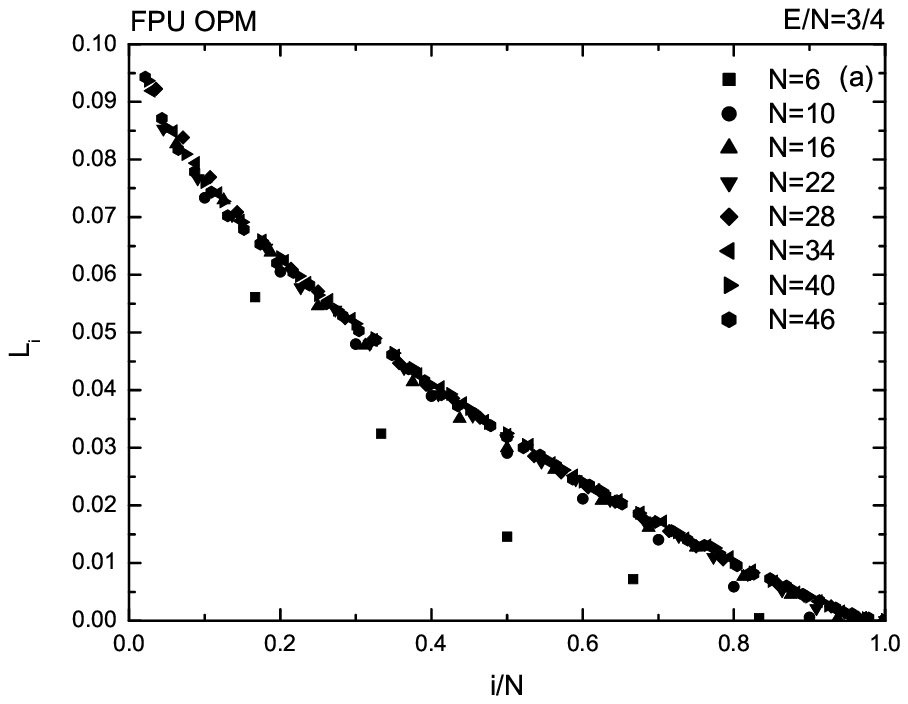} \includegraphics{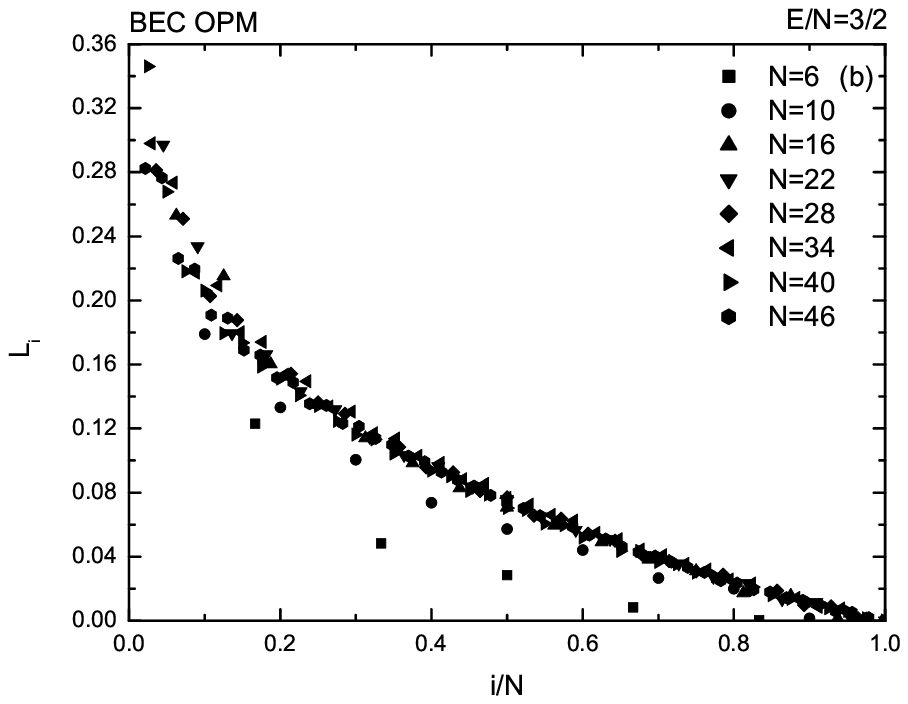} \vspace{4cm}
\end{centering}
\caption{(a) Positive Lyapunov exponents spectrum of the OPM
(\ref{FPU_non_lin_mode_periodic_boundary_conditions_OPM}) of the FPU
Hamiltonian (\ref{FPU_Hamiltonian_2}) for fixed
$\frac{E}{N}=\frac{3}{4}$. (b) Positive Lyapunov exponents spectrum
of the OPM (\ref{BEC_non_lin_mode_periodic_boundary_conditions_OPM})
of the BEC Hamiltonian (\ref{BEC_Hamiltonian_2}) for fixed
$\frac{E}{N}=\frac{3}{2}$. In both panels $i$ runs from $1$ to $N$.}
\label{fig_9}
\end{figure}

Finally, choosing again as initial conditions the unstable OPMs of
both Hamiltonians, we determine some important statistical
properties of the dynamics in the so--called thermodynamic limit of
$E$ and $N$ growing indefinitely, while keeping $E/N$ constant. In
particular, we compute the spectrum of the Lyapunov exponents of the
FPU and BEC systems starting at the OPM solutions
(\ref{FPU_non_lin_mode_periodic_boundary_conditions_OPM}) and
(\ref{BEC_non_lin_mode_periodic_boundary_conditions_OPM}) for
energies where these orbits are unstable. We thus find that the
Lyapunov exponents are well approximated by smooth curves of the
form $L_{i}\approx L_1 e^{-\alpha i/N}$, for both systems, with
$\alpha\approx 2.76$, $\alpha\approx 3.33$ respectively and
$i=1,2,...,K(N)$ where $K(N)\approx3N/4$ (see Fig.~\ref{fig_9}).

Specifically, in the case of the OPM
(\ref{FPU_non_lin_mode_periodic_boundary_conditions_OPM}) of the FPU
Hamiltonian (\ref{FPU_Hamiltonian_2}) for fixed energy density
$\frac{E}{N}=\frac{3}{4}$ we find
\begin{equation}\label{L_i_FPU_OPM}
L_{i}(N)\approx L_1(N) e^{-2.76\frac{i}{N}},
\end{equation}
while, in the case of the OPM
(\ref{BEC_non_lin_mode_periodic_boundary_conditions_OPM}) of the
BEC Hamiltonian (\ref{BEC_Hamiltonian_2}) for fixed energy density
$\frac{E}{N}=\frac{3}{2}$, a similar behavior is observed,
\begin{equation}\label{L_i_BEC_OPM}
L_{i}(N)\approx L_1(N) e^{-3.33\frac{i}{N}} .
\end{equation}
These exponential formulas were found to hold quite well, up to
$i=K(N)\approx3N/4$. For the remaining exponents, the spectrum is
seen to obey different decay laws, which are not easy to
determine. As this appears to be a subtle matter, however, we
prefer to postpone it for a future publication.

The functions (\ref{L_i_FPU_OPM}) and (\ref{L_i_BEC_OPM}), provide
in fact, invariants of the dynamics, in the sense that, in the
thermodynamic limit, we can use them to evaluate the average of the
positive Lyapunov exponents (i.e. the Kolmogorov--Sinai entropy per
particle) for each system and find that it is a constant
characterized by the value of the maximum Lyapunov exponent $L_1$
and the exponent $\alpha$ appearing in them.

\begin{figure}[ht]
\begin{center}
\includegraphics{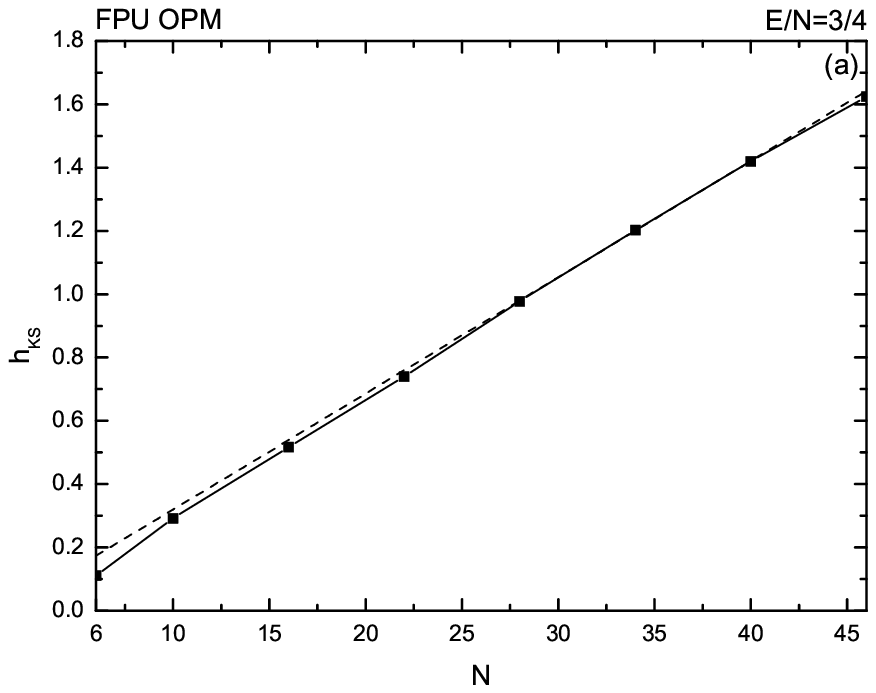} \includegraphics{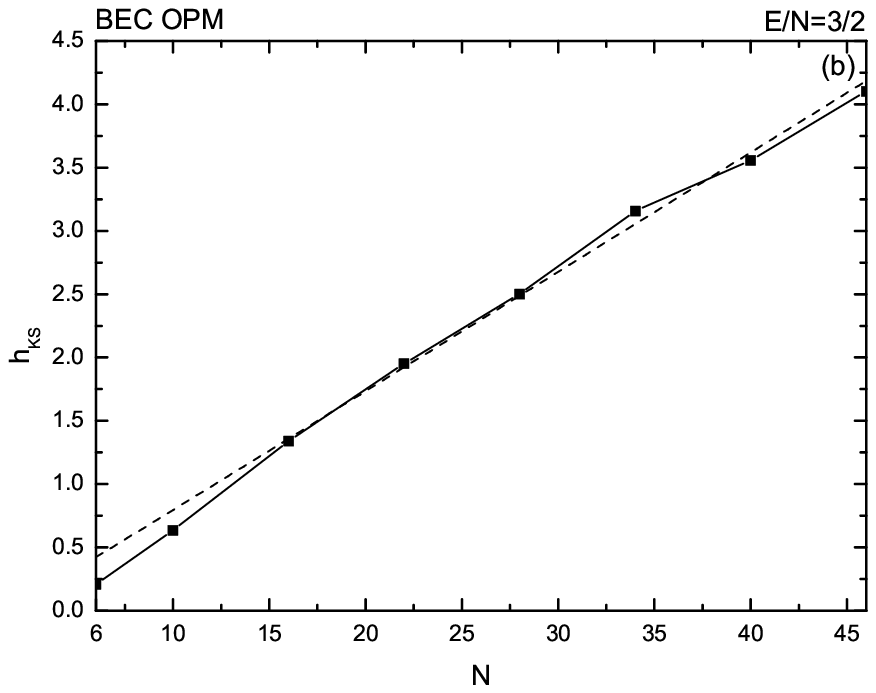} \vspace{3.3cm}
\end{center}
\caption{(a) The $h_{KS}(N)$ entropy of the OPM
(\ref{FPU_non_lin_mode_periodic_boundary_conditions_OPM}) of the
FPU Hamiltonian (\ref{FPU_Hamiltonian_2}) for fixed
$\frac{E}{N}=\frac{3}{4}$ (solid curve) and the approximated
formula (\ref{h_KS_FPU_OPM}) (dashed curve). (b) The $h_{KS}(N)$
entropy of the OPM
(\ref{BEC_non_lin_mode_periodic_boundary_conditions_OPM}) of the
BEC Hamiltonian (\ref{BEC_Hamiltonian_2}) for fixed
$\frac{E}{N}=\frac{3}{2}$ (solid curve) and the approximated
formula (\ref{h_KS_BEC_OPM}) (dashed curve).} \label{graph_16}
\end{figure}

In Fig.~\ref{graph_16} we compute the well--known Kolmogorov--Sinai
entropy $h_{KS}(N)$ \cite{Pesin,Hilborn} (solid curves), which is
defined as the sum of the $N-1$ positive Lyapunov exponents,
\begin{equation}\label{KS_entropy}
h_{KS}(N)=\sum_{i=1}^{N-1}L_{i}(N),\ L_{i}(N)>0.
\end{equation}
In this way, we find, for both Hamiltonians, that $h_{KS}(N)$ is
an extensive thermodynamic quantity as it is clearly seen to grow
linearly with $N$ ($h_{KS}(N)\propto N$), demonstrating that in
their chaotic regions the FPU and BEC Hamiltonians behave as
ergodic systems of statistical mechanics.

Finally, using Eqs.~(\ref{L_i_FPU_OPM}) and (\ref{L_i_BEC_OPM}),
as if they were valid for all $i=1,\ldots,N-1$, we approximate the
sum of the positive Lyapunov exponents $L_i$, and calculate the
$h_{KS}(N)$ entropy from Eq.~(\ref{KS_entropy}) as
\begin{equation}\label{h_KS_FPU_OPM}
h_{KS}(N)\propto
L_{\mathrm{max}}\frac{1}{-1+e^{\frac{2.76}{N}}},\,\,\,
L_{\mathrm{max}}\approx 0.095, \mathrm{(FPU\ OPM)}
\end{equation}
and
\begin{equation}\label{h_KS_BEC_OPM}
h_{KS}(N)\propto
L_{\mathrm{max}}\frac{1}{-1+e^{\frac{3.33}{N}}},\,\,\,
L_{\mathrm{max}}\approx 0.34, \mathrm{(BEC\ OPM)}.
\end{equation}
In Fig.~\ref{graph_16}, we have plotted Eqs.~(\ref{h_KS_FPU_OPM})
and (\ref{h_KS_BEC_OPM}) with dashed curves (adjusting the
proportionality constants appropriately) and obtain nearly
straight lines with the same slope as the data computed by the
numerical evaluation of the $h_{KS}(N)$, from
Eq.~(\ref{KS_entropy}).

\section{Conclusions}\label{concl_6}
In this paper we have investigated the connection between local and
global dynamics of two $N$ dof Hamiltonian systems, describing $1$D
nonlinear lattices with different origins known as the FPU and BEC
systems. We focused on solutions located in the neighborhood of
simple periodic orbits and showed that as the energy increases
beyond the destabilization threshold, all positive Lyapunov
exponents increase monotonically with two distinct power--law
dependencies on the energy. We also computed the destabilization
energy per particle of the OHS mode and of the two OPM orbits and
found that it decays with a simple power--law of the form
$E_c/N\propto N^{-\alpha}$, $\alpha=$ 1 or 2. One notable exception
is the IPM orbit of the BEC Hamiltonian, which is found to be stable
for any energy and number of dof we considered!

Furthermore, we found that as we increase the energy $E$ of both
Hamiltonians for fixed $N$, the SPOs behave in very different ways:
In the OPM case of the FPU Hamiltonian the eigenvalues of the
monodromy matrix always move away from $-1$ with increasing energy,
while in the OPM of the BEC Hamiltonian these eigenvalues return to
$+1$, for very high energies. Nevertheless, this behavior does not
represent a return of the system to a globally regular motion around
the SPO as one might have thought. It simply reflects a local
property of the SPO, as the orbits in its neighborhood still have
positive Lyapunov exponents which are far from zero.

We have also been able to estimate the ``size'' of the islands of
stability around the SPOs with the help of a recently introduced
very efficient indicator called SALI and have seen them to shrink,
as expected, when increasing $N$ for fixed $E$ in the case of the
OPM of the FPU Hamiltonian. Of course, when we continue increasing
the energy, keeping the number of dof fixed, the OPMs destabilize
and the islands of stability are destroyed. Unexpectedly, however,
in the case of the OPM of the BEC Hamiltonian these islands were
seen to {\it grow} in size if we keep $N$ fixed and increase $E$ up
to the destabilization threshold, a result that clearly requires
further investigation. The peculiarity of our BEC Hamiltonian is
even more vividly manifested in the fact that the islands of
stability about its IPM orbit never vanish, remaining actually of
significant size, even for very high values of $E$ and $N$ (keeping
$E/N$ fixed).

Starting always near unstable SPOs, we also calculated the Lyapunov
spectra characterizing chaotic dynamics in the regions of the OHS
and OPM solutions of both Hamiltonians. Keeping $N$ fixed, we found
energy values where these spectra were practically the same near
SPOs which are far apart in phase space, indicating that the chaotic
regions around these SPOs are visited ergodically by each other's
orbits. Finally, using an exponential law accurately describing
these spectra, we were able to show, for both Hamiltonians, that the
associated Kolmogorov--Sinai entropies per particle increase
linearly with $N$ in the thermodynamic limit of $E\rightarrow\infty$
and $N\rightarrow\infty$ and fixed $E/N$ and, therefore, behave as
extensive quantities of statistical mechanics.

Our results suggest, however, that, even in that limit, there may
well exist Hamiltonian systems with significantly sized islands of
stability around stable SPOs (like the IPM of the BEC Hamiltonian),
which must be excluded from a rigorous statistical description. It
is possible, of course, that these islands are too small in
comparison with the extent of the chaotic domain on a constant
energy surface and their ``measure'' may indeed go to zero in the
thermodynamic limit. But as long as $E$ and $N$ are finite, they
will still be there, precluding the global definition of probability
densities, ensemble averages and the validity of the ergodic
hypothesis over all phase space. Clearly, therefore, their study is
of great interest and their properties worth pursuing in Hamiltonian
systems of interest to physical applications.


\section{Acknowledgements}
This work was partially supported by the European Social Fund (ESF),
Operational Program for Educational and Vocational Training II
(EPEAEK II) and particularly the Program HERAKLEITOS, providing a
Ph.~D scholarship for one of us (C.~A.). C.~A. also acknowledges
with gratitude the $3$ month hospitality, March--June $2005$, of the
``Center for Nonlinear Phenomena and Complex Systems'' of the
University of Brussels. In particular, he thanks Professor
G.~Nicolis, Professor P.~Gaspard and Dr.~V.~Basios for their
instructive comments and useful remarks in many discussions
explaining some fundamental concepts treated in this paper. The
second author (T.~B.) wishes to express his gratitude to the Max
Planck Institute of the Physics of Complex Systems at Dresden, for
its hospitality during his $3$ month visit March--June $2005$, when
this work was completed. In particular, T.~B.~wants to thank
Dr.~Sergej Flach for numerous lively conversations and exciting
arguments on the stability of multi--dimensional Hamiltonian
systems. Useful discussions with Professors A.~Politi, R.~Livi,
R.~Dvorak, F.~M.~Izrailev and Dr.~T.~Kottos are also gratefully
acknowledged. The third author (C.~S.) was partially supported by
the Research Committee of the Academy of Athens.


\section{Figure Captions}

\begin{enumerate}
\item{The solid curve corresponds to the energy per particle
$\frac{E_{c}}{N}$, for $\beta=1.04$, of the first destabilization of
the OHS nonlinear mode
(\ref{FPU_non_lin_mode_fixed_boundary_conditions_OHS}) of the FPU
system (\ref{FPU_Hamiltonian_2}) obtained by the numerical
evaluation of the Hill's determinant in
(\ref{FPU_stability_criterion_OHS}), while the dashed line
corresponds to the function $\propto\frac{1}{N}$. Note that both
axes are logarithmic.}

\item{The solid curve corresponds to the energy per particle
$\frac{E_{c}}{N}$, for $\beta=1$, of the first destabilization of
the nonlinear OPM
(\ref{FPU_non_lin_mode_periodic_boundary_conditions_OPM}) of the FPU
system (\ref{FPU_Hamiltonian_2}) obtained by the numerical
evaluation of the Hill's determinant, while the dashed line
corresponds to the function $\propto\frac{1}{N^2}$. Note that both
axes are logarithmic.}

\item{The solid curve corresponds to the energy per particle
$\frac{E_{c}}{N}$ of the first destabilization of the OPM
(\ref{BEC_non_lin_mode_periodic_boundary_conditions_OPM}) of the BEC
Hamiltonian (\ref{BEC_Hamiltonian_2}) obtained by the numerical
evaluation of the eigenvalues of the monodromy matrix of Eq.
(\ref{BEC_linearized_equations}), while the dashed line corresponds
to the function $\propto\frac{1}{N^{2}}$. Note that both axes are
logarithmic.}

\item{The two distinct power--law behaviors in the evolution of
the maximum Lyapunov exponent $L_1$ as the energy grows for the OPM
(\ref{FPU_non_lin_mode_periodic_boundary_conditions_OPM}) of the FPU
Hamiltonian (\ref{FPU_Hamiltonian_2}) for $N=10$. A similar picture
is obtained for the $L_2$ and $L_3$ also, with similar exponents and
the transition occurring at the same energy value (see text). Note
that both axes are logarithmic.}

\item{(a) The spectrum of the positive Lyapunov exponents for
fixed $N=10$ of the OPM
(\ref{FPU_non_lin_mode_periodic_boundary_conditions_OPM}) of the FPU
Hamiltonian (\ref{FPU_Hamiltonian_2}) as the energy grows. (b) Also
for $N=10$, the OPM
(\ref{BEC_non_lin_mode_periodic_boundary_conditions_OPM}) of the BEC
Hamiltonian (\ref{BEC_Hamiltonian_2}) yields a similar picture as
the energy is increased. (c) The Lyapunov spectrum of the OPM
(\ref{FPU_non_lin_mode_periodic_boundary_conditions_OPM}) of the FPU
Hamiltonian (\ref{FPU_Hamiltonian_2}) for $N=16$ and the OHS mode
(\ref{FPU_non_lin_mode_fixed_boundary_conditions_OHS}) of the same
Hamiltonian and $N$, for periodic boundary conditions practically
coincide at $E=6.82$ where both of them are destabilized. (d) The
Lyapunov spectrum of the FPU OHS mode
(\ref{FPU_non_lin_mode_fixed_boundary_conditions_OHS}) with fixed
boundary conditions for $N=15$ as the energy grows presents as shape
which is qualitatively similar to what was found for the SPOs of
panel (c).}

\item{The eigenvalues $\lambda_{j},\;j=1,\ldots,2N$ (a) of the
OPM (\ref{FPU_non_lin_mode_periodic_boundary_conditions_OPM}) of the
FPU Hamiltonian (\ref{FPU_Hamiltonian_2}) with $N=10$ dof. (b) The
eigenvalues of the OPM
(\ref{BEC_non_lin_mode_periodic_boundary_conditions_OPM}) of the BEC
Hamiltonian (\ref{BEC_Hamiltonian_2}) with the same number of dof
$N$. (c) The eigenvalues of the IPM
(\ref{BEC_non_lin_mode_periodic_boundary_conditions_IPM}) of the BEC
Hamiltonian (\ref{BEC_Hamiltonian_2}) with $N=6$ dof.}

\item{(a) The time evolution of the SALI (solid curve) and of the
Eq.~(\ref{SALI_MLE_relation}) (dashed line) at $E=21.6$ of the OHS
mode (\ref{FPU_non_lin_mode_fixed_boundary_conditions_OHS}) of the
FPU Hamiltonian (\ref{FPU_Hamiltonian_2}) with fixed boundary
conditions. (b) Similar plot to panel (a) but for the larger energy
$E=26.6878$ for which the two largest Lyapunov exponents $L_1$ and
$L_2$ are almost equal while all the other positive ones are very
close to zero. In both panels the agreement between the data (solid
curve) and the derived function of Eq.~(\ref{SALI_MLE_relation})
(dashed line) is remarkably good. Note that the horizontal axes in
both panels are linear.}

\item{(a) ``Size'' of the islands of stability of the OPM
(\ref{BEC_non_lin_mode_periodic_boundary_conditions_OPM}) of the BEC
Hamiltonian (\ref{BEC_Hamiltonian_2}) for $N=4,\ 6$ and $8$ dof and
SPOs constant energy $E=1$ before the first destabilization (see
Table \ref{table_BEC_OPM}). (b) ``Size'' of the islands of stability
of the same Hamiltonian and SPO as in (a) for $N=6$ dof and three
different energies of the SPO before the first destabilization (see
Table \ref{table_BEC_OPM}). (c) ``Size'' of the islands of stability
of the same Hamiltonian as in (a) of the IPM
(\ref{BEC_non_lin_mode_periodic_boundary_conditions_IPM}) for $N=6$
dof and four different energies of the SPO. (d) ``Size'' of the
islands of stability of the same Hamiltonian as in (a) of the IPM
(\ref{BEC_non_lin_mode_periodic_boundary_conditions_IPM}) for
$\frac{E}{N}=\frac{10}{3}$. Here $E$ corresponds to the energy of
the IPM.}

\item{(a) Positive Lyapunov exponents spectrum of the OPM
(\ref{FPU_non_lin_mode_periodic_boundary_conditions_OPM}) of the FPU
Hamiltonian (\ref{FPU_Hamiltonian_2}) for fixed
$\frac{E}{N}=\frac{3}{4}$. (b) Positive Lyapunov exponents spectrum
of the OPM (\ref{BEC_non_lin_mode_periodic_boundary_conditions_OPM})
of the BEC Hamiltonian (\ref{BEC_Hamiltonian_2}) for fixed
$\frac{E}{N}=\frac{3}{2}$. In both panels $i$ runs from $1$ to $N$.}

\item{(a) The $h_{KS}(N)$ entropy of the OPM
(\ref{FPU_non_lin_mode_periodic_boundary_conditions_OPM}) of the
FPU Hamiltonian (\ref{FPU_Hamiltonian_2}) for fixed
$\frac{E}{N}=\frac{3}{4}$ (solid curve) and the approximated
formula (\ref{h_KS_FPU_OPM}) (dashed curve). (b) The $h_{KS}(N)$
entropy of the OPM
(\ref{BEC_non_lin_mode_periodic_boundary_conditions_OPM}) of the
BEC Hamiltonian (\ref{BEC_Hamiltonian_2}) for fixed
$\frac{E}{N}=\frac{3}{2}$ (solid curve) and the approximated
formula (\ref{h_KS_BEC_OPM}) (dashed curve).}

\end{enumerate}
\end{document}